\documentclass[aps,prx,reprint,showpacs,superscriptaddress,citeautoscript,footinbib]{revtex4-1}

\usepackage{color}
\usepackage{graphicx}
\usepackage{dcolumn}
\usepackage{bm}
\usepackage{url}
\usepackage[colorlinks=True,citecolor=blue,linkcolor=blue,urlcolor=blue]{hyperref}
\usepackage{amssymb, amsmath}

\begin{document}



\title{Lattice dynamics in FeSe via inelastic x-ray scattering and first-principles calculations}


\author{Naoki Murai}
\email{naoki.murai@j-parc.jp}
\affiliation{Materials and Life Science Division, J-PARC Center, Japan Atomic Energy Agency, Tokai, Ibaraki 319-1195, Japan\looseness=-1}
\affiliation{Materials Dynamics Laboratory, RIKEN SPring-8 Center, 1-1-1 Kouto, Sayo, Hyogo 679-5148, Japan}
\author{Tatsuo Fukuda}
\affiliation{Materials Sciences Research Center, Japan Atomic Energy Agency (SPring-8/JAEA), 1-1-1 Kouto, Sayo, Hyogo 679-5148, Japan}
\affiliation{Materials Dynamics Laboratory, RIKEN SPring-8 Center, 1-1-1 Kouto, Sayo, Hyogo 679-5148, Japan}
\author{Masamichi Nakajima}
\affiliation{Department of Physics, Osaka University, Toyonaka,Osaka 560-0043, Japan}
\author{Mitsuaki~Kawamura}
\affiliation{Institute for Solid State Physics, The University of Tokyo, Kashiwa 277-8581, Japan}
\author{Daisuke Ishikawa}
\affiliation{Research and Utilization Division, Japan Synchrotron Radiation Research Institute (SPring-8/JASRI),
  1-1-1 Kouto, Sayo, Hyogo 679-5198, Japan\looseness=-1}
\affiliation{Materials Dynamics Laboratory, RIKEN SPring-8 Center, 1-1-1 Kouto, Sayo, Hyogo 679-5148, Japan}
\author{Setsuko Tajima}
\affiliation{Department of Physics, Osaka University, Toyonaka,Osaka 560-0043, Japan}
\author{Alfred Q. R. Baron}
\affiliation{Materials Dynamics Laboratory, RIKEN SPring-8 Center, 1-1-1 Kouto, Sayo, Hyogo 679-5148, Japan}


\date{\today}
\begin{abstract}
  We report an inelastic x-ray scattering investigation of phonons in FeSe superconductor.
  Comparing the experimental phonon dispersion with density functional theory (DFT)
  calculations in the non-magnetic state, we found a significant disagreement between them. 
  Improved overall agreement was obtained by allowing for spin-polarization in the DFT calculations, 
  despite the absence of magnetic order in the experiment.
  This calculation gives a realistic approximation, at DFT level, of the disordered paramagnetic
  state of FeSe, in which strong spin fluctuations are present. 
\end{abstract}

\maketitle
\section{Introduction}
\indent The surprising discovery of high-\(T_{\rm c}\) superconductivity in iron-based superconductors (FeSCs)\cite{Kamihara2008JACS}
has marked the beginning of a new era in superconductivity research. Many of the properties in FeSCs arise from a coupling of
spin, orbital and lattice degrees of freedom. 
In particular, the interplay of structure and magnetism, often referred to as magneto-elastic coupling,
is one of the most engaging topics in the study of FeSCs, as it is increasingly recognized that
these two degrees of freedom significantly impact each other.
The early evidence of this relationship was the observation of the collapsed tetragonal phases\cite{Kreyssig2008Phys.Rev.B},
in which the Fe-magnetism has been shown to have a significant impact on the crystal structure. 
Subsequently, after it was demonstrated that phonon calculations without magnetic order failed to match the
measured dispersion\cite{Fukuda2008JPSJ}, similar phenomena have also been observed in phonon dispersion
measurements of FeSCs\cite{Hahn2009PhysRevB,Reznik2009Phys.Rev.B,Fukuda2011Phhys.Rev.B}.
More recently, we have demonstrated that
the presence of the stripe-type antiferromagnetic (AFM) order lifts the degeneracy of phonon frequencies, which results in
a symmetry-breaking modification of the overall phonon structure\cite{Murai2016Phys.Rev.B}.
These results highlight the unprecedented sensitivity of the lattice dynamics to the underlying magnetic structure.\\
\indent So far, while the phonon measurements and calculations in FeSCs have been performed with an emphasis on the role of the static long-range
magnetic order, those in the paramagnetic phase remain relatively unexplored. Indeed, the effects of melting magnetic order with temperature, and
the impact of the resulting disordered magnetism on phonons were reported only recently\cite{Murai2016Phys.Rev.B}. 
It is therefore of particular interest to investigate the lattice dynamics of the disordered paramagnetic state,
in which the static magnetic order is replaced by local spin fluctuations.\\
\indent FeSe, the structurally simplest FeSCs, provides an excellent platform for studying such issues, because, in contrast with the other
FeSCs, no static magnetic order occurs down to the lowest temperature\cite{Bohmer2017JPCM}. 
Here we report the results of meV-resolved inelastic x-ray scattering (IXS) measurements on FeSe superconductor.
We found that the experimental phonon dispersion of FeSe deviates significantly from the prediction of density functional theory (DFT)
calculations in the non-magnetic (NM) state. 
A better overall agreement is obtained by allowing for full spin-polarization in the DFT calculations,
despite the absence of magnetic order at ambient pressure. 
The present results show that the inclusion of magnetism within DFT is crucial to reproduce the lattice dynamics of the disordered
paramagnetic state located near the magnetic instability.
\section{Experimental and computational details}
\begin{figure}[tb]
\begin{center}
  \includegraphics[clip,width= 8.30000cm]{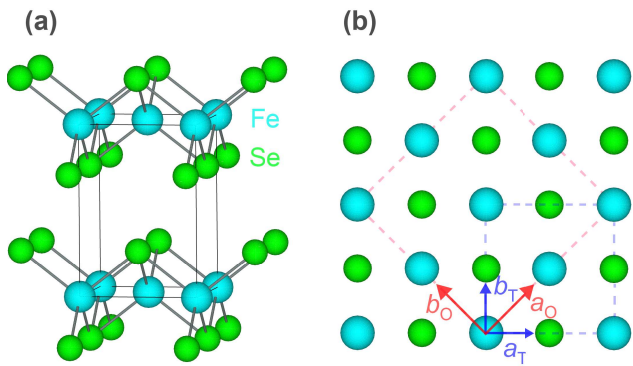}
  \caption{
    (Color online) Crystal structure and unit cell conventions of FeSe. (a) Crystal structure of FeSe in the tetragonal
    phase ($\it P{\rm 4}/nmm$). (b) Top view of the crystal structure. Dashed blue and red squares indicate the tetragonal
    and orthorhombic unit cells, respectively. 
    The tetragonal lattice parameters are related to the orthorhombic one by
    \(a_{\rm 0} \sim b_{\rm 0} = \sqrt{2}a_{\rm T}\).
    The blue and red arrows indicate the directions of tetragonal and orthorhombic lattice vectors, respectively.
    The crystal structures were visualized using the {\scshape{vesta}} software\cite{Momma2011}.
  }
\label{Fig1}
\end{center}
\end{figure}

\indent Single crystals of FeSe were grown by a chemical vapor transport method similar to that described in
Ref.[\onlinecite{Bohmer2016PhysRevB.94.024526}]. Fe and Se powders with a molar ratio of 1.2:1 were sealed in an evacuated quartz tube together with a
eutectic mixture of KCl and AlCl$_{\rm 3}$ as a transport agent.
The quartz tube was placed in a tilted tube furnace and heated at 350 $^\circ\mathrm{C}$ and
390 $^\circ\mathrm{C}$ for the sealed and the other end, respectively.
After 20-30 days, millimeter-sized single crystals were obtained in the cold end.
Upon cooling, FeSe exhibits a structural phase transition from a tetragonal (\(P4/nmm\)) to an orthorhombic (\(Cmma\)) crystal symmetry 
at \(T_{\rm s} \sim {\rm 90}~ {\rm K}\). (The crystal structure of FeSe and its unit cell conventions are schematically displayed in Fig. \ref{Fig1}.)
Throughout this paper, we define the momentum transfer \({\bm Q} = H{\bm a^{*}} + K{\bm b^{*}} + L{\bm c^{*}} \equiv (H,K,L)\)
in reciprocal lattice units (r.l.u.) by using the tetragonal unit cell.\\
\indent High-resolution IXS measurements of FeSe were performed at BL43LXU\cite{Baron2010riken,Baron2016} of the SPring-8 in Japan.
An incident x-ray energy of 21.747~keV, which corresponds to Si(11 11 11) reflection, gives an energy resolution of about 1.5~meV,
depending on the analyzer crystals.
The use of a two-dimensional (2D) analyzer array allowed for the parallelization of data collection in a
2D section of momentum space\cite{Baron2016, Baron2008JPCM}.
To extract the phonon dispersion of FeSe, IXS spectra were fitted to the sum of a resolution-limited
elastic peak and several damped harmonic oscillators for the phonon modes convoluted with the experimentally determined
resolution functions. The best fit parameters and their errors were obtained using the {\scshape{minuit}}
minimization code\cite{James1975Comput.Phys.Commun.} in the CERN program library. \\
\indent To understand the lattice dynamics of FeSe, we performed first-principles phonon calculations
using the density-functional perturbation theory (DFPT)\cite{Baroni2001Rev.Mod.Phys},
as implemented in {\scshape{quantum espresso}} code
\cite{Giannozzi2009J.Phys:Condensed.Matter, Giannozzi2017J.Phys:Condensed.Matter}.
In all calculations, the exchange correlation functional was treated within the generalized gradient approximation (GGA) using 
the Perdew-Burke-Ernzerhof parameterization for solids (PBEsol)\cite{Perdew2008Phys.Rev.Lett}.
We used an ultrasoft pseudopotentials from pslibrary\cite{DalCorso2014Comput.Mater.Sci.} 
with cutoffs of 90 and 1080 Ry for the expansion of the wave functions and charge densities, respectively.
We also performed the calculation by setting these cutoffs to 100 Ry and 1200 Ry to check the convergence of
phonon frequencies with respect to the number of plane waves. 
The Brillouin zone integration was performed over a \(12\times12\times12\) \(\bm k\) mesh with
a smearing of 0.01 Ry.
The lattice parameters were fixed to the experimental values\cite{Matsuura2017Nat.Commun} and the internal parameter
(i.e., the Se position) was optimized. 
Dynamical matrices were calculated on \(4\times4\times4\) uniform grids in \(\bm q\) space, which were then
interpolated to determine the full phonon dispersion.
All calculations were performed for both NM and AFM states. 
In the former case, the occupation numbers of spin-up and spin-down Fe-3$d$ states are forced to be equal
(i.e., the magnetic moment is constrained to be zero), 
while those in the latter are allowed to vary independently
(i.e., the Fe atoms are allowed to have a nonzero magnetic moment corresponding to the minimum in total energy).
\begin{figure}[t]
\begin{center}
  \includegraphics[clip,width= 7.80000cm]{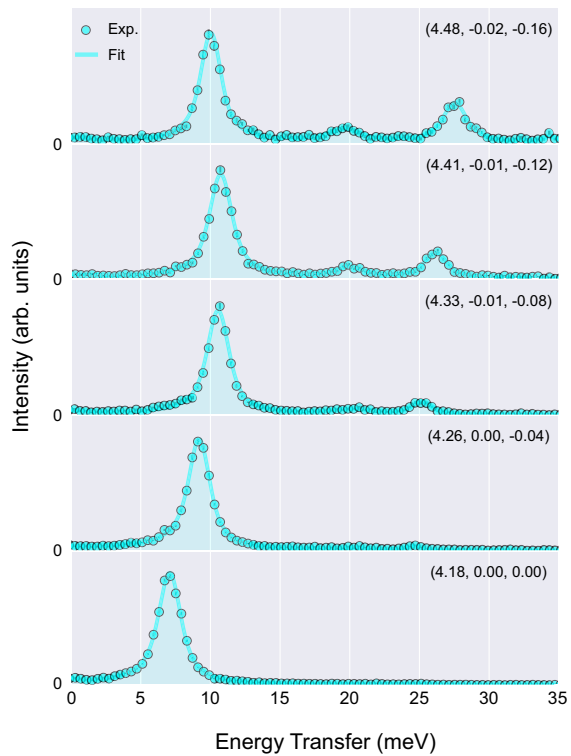}
  \caption{
    (Color online) Momentum dependence of the IXS spectra along the \([1 \ 0 \ 0]\) direction
    in the \((4, 0, 0)\) Brillouin zone. Data were collected at 150 K.  
    The experimental data (circles) are shown together with the best fits to the data
    (solid lines). }
\label{Fig2}
\end{center}
\end{figure}
\begin{figure*}[tht]
\begin{center}
  \includegraphics[clip,width= 17.50000cm]{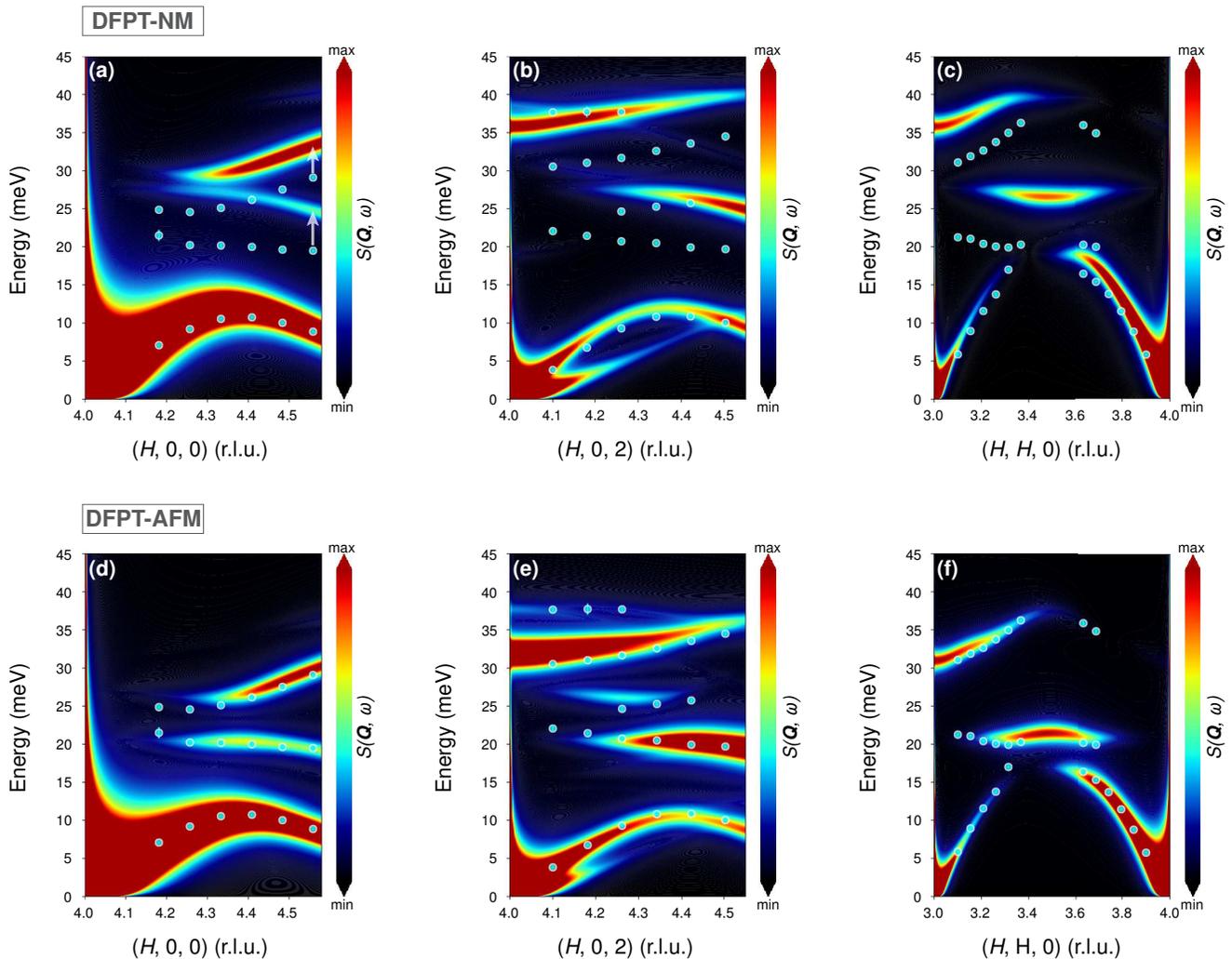}
  \caption{
    (Color online) Comparison of the measured phonon dispersion for FeSe and DFPT calculations in 
    the NM ((a)-(c)) and the AFM ((d)-(f)) cases.
    The markers are experimental data extracted from fits to the IXS spectra at 150 K. 
    To facilitate the comparison between theory and experiment,
    the dynamical structure factor, \(S({\bm Q}, \omega)\), is weighted on the calculated phonon dispersion curves.
    Arrows in Fig. \ref{Fig3} (a) are guides to the eyes showing the discrepancy between the experimental and calculated
      optical branches. 
    }
\label{Fig3}
\end{center}
\end{figure*}
\section{Results and discussion}
\indent Figure \ref{Fig2} shows representative IXS spectra of FeSe along the [1 0 0] direction.
Overall, no obvious anomalies in the phonon modes (such as large linewidth or anomalous dispersion) were found.  
The same conclusions can also be obtained from the IXS scans along other high symmetry directions. 
Conversely, as we will show below, there is a significant disagreement between the measured and the calculated phonon dispersion. \\
\indent Figures \ref{Fig3} (a)-(c) show the observed phonon dispersions (cyan circles) and 
their comparison with DFT calculations in the NM state 
   \footnote{
      Due to the finite momentum acceptance of the analyzers, some unexpected modes appear in the measured spectra
      (e.g, transverse modes in the longitudinal directions). To facilitate an easy comparison between theory and experiment,
      we here plot the experimental data expected in the measured symmetry directions.
   }.
To facilitate the comparison between theory and experiment, 
the dynamical structure factor, \(S({\bm Q}, \omega)\)\cite{Baron2016,squires_2012}, is weighted on the DFT-calculated dispersion curves.
The DFT calculations are reasonably consistent with the acoustic branches; however, in most parts of the
Brillouin zone investigated, they fail to reproduce the optical branches (see arrows in Fig. \ref{Fig3} (a)).
Overall, the experimental optical modes are much softer (about 5~meV) than those calculated. 
Similar behaviors have been found in other FeSCs\cite{Fukuda2008JPSJ,Hahn2009PhysRevB,Reznik2009Phys.Rev.B,Fukuda2011Phhys.Rev.B,Murai2016Phys.Rev.B,Baron2016}. \\
\indent In general, DFT calculations are remarkably successful in predicting some of the relevant properties of a wide range of FeSCs,
such as the Fermi surface topology\cite{Watson2019Quantum_Materials,Fedorov2019PhysRevB,2019PtoKPhysRevB}
and the stable magnetic structure\cite{Yildirim2008PhysRevLett,Ishibashi2008JPSJ}. However, it is well known that they fail to reproduce the
correct structural \cite{Mazin2008Nat.Phys.} and vibrational\cite{Fukuda2008JPSJ} properties of FeSCs.
This limitation is mainly due to the presence of local magnetic moments in the paramagnetic state,
which have a significant impact on structural properties.
The importance of this effect was suggested early on in Ref. [\onlinecite{Mazin2008Nat.Phys.}], but the first-principles
treatment of the disordered paramagnetic state remains computationally challenging\cite{Abrikosov2016}.
At ambient pressure, FeSe does not order magnetically, but, as in the case of other FeSCs,
the stripe-type AFM spin fluctuations are clearly observed at finite energy transfers
\cite{Boothroyd2015Phys.RevB, Wang2015Nat.Mater, Wang2016Nat.Commun., Ma2017Phys.Rev.X, Chen2019Nat.Mater}.
The failure of the non-magnetic DFT approach is, therefore, not surprising considering the importance
of dynamically fluctuating spin correlations in FeSe. 
A relatively simple way to account for such an effect is to use spin-polarized DFT calculations. 
One should, however, note that this approach assumes some sort of static magnetic order,
so the effect of magnetism is computed only on the static DFT level.
Nevertheless, as detailed below, spin-polarized DFT calculations give a reasonable approximation to the structural properties 
in the disordered paramagnetic phase.  \\
\indent To highlight the sensitivity of the structural properties to magnetism,
we summarize in Table \ref{table:structure} the results of the calculated 
Se position {\it z$_{ \scalebox{0.55}{\rm { }Se}}$} with and without the AFM order. 
For the spin configuration of the AFM state, we consider the stripe pattern, in which spins are aligned (anti)ferromagnetically
along the tetragonal [1 1 0] (\([1 \ \overline{1} \ 0]\)) directions. 
The inclusion of the magnetism within spin-polarized DFT has a significant effect on {\it z$_{ \scalebox{0.55}{\rm { }Se}}$},  
which improves the agreement with experiment. As a related point, we note that 
an orthorhombic structure of FeSe quantitatively consistent with the experiment can only be obtained when
structurally optimized in magnetically ordered states\cite{Glasbrenner2015Nat.Pys.}. \\
\begin{table}[tb]
  \caption{Chalcogen position parameter \(z_{\rm se}\) for FeSe computed by the non-magnetic DFT (DFT-NM) and
    the spin-polarized DFT in the AFM state (DFT-AFM).
    Here, the Se positions were optimized while keeping lattice parameters fixed to those experimentally determined\cite{Matsuura2017Nat.Commun}. 
    Note that while the optimized values of $z_{{\rm { } se}}$ are sensitive to the choice of
    exchange-correlation functional and pseudopotential, 
    the structural optimization with magnetism
    generally agrees better with the experiment.
  }
  \label{table:structure}
  \centering
  \begin{tabular}{lccc}
    \hline
    \hline
      & { }{ }Exp\cite{Matsuura2017Nat.Commun}.{ }{ } & { } DFT-NM { } & { } DFT-AFM { } \\
    \hline
     $z_{{\rm { } se}}$  & 0.267  & 0.235 & 0.251 \\
     \hline
    \hline
 \end{tabular}
\end{table}
\indent Interestingly, similar improvements can also be seen in phonon dispersions.
In Figs. \ref{Fig3} (d) - (f), the experimental phonon data are compared with spin-polarized DFPT phonon calculations\footnote{
  There are two inequivalent [1 1 0] directions in the spin-polarized DFPT phonon calculation, but 
  the anisotropy is not so pronounced between these two direction. 
  Hence, 
  the calculated phonon dispersion is shown along one particular [1 1 0] direction 
  }.
The inclusion of magnetism clearly results in a frequency shift of some modes 
and brings them into much better agreement with the experimental data. 
The AFM order has the biggest effect on high-energy optical branches, with
lower energy acoustic branches relatively unaffected. 
These results are reminiscent of other FeSCs, in which lattice dynamics properties are very sensitive to the
underlying magnetic state\cite{Hahn2009PhysRevB,Reznik2009Phys.Rev.B,Fukuda2011Phhys.Rev.B,Murai2016Phys.Rev.B,Baron2016}. 
One should note that phonon calculations are usually preformed for the DFT-optimized structure, and thus the resulting phonon dispersion
is affected not only by the magnetic ground state but also by the structural details. 
To distinguish these two effects, we performed the non-magnetic DFPT phonon calculation using the
crystal structure optimized in the AFM state.
This calculation slightly improves consistency with the experiments, but not so much as the spin-polarized DFPT calculation.
(see \ref{appendix}. Appendix for details.)
The improved agreement shown in Figs. \ref{Fig3} (d) -(f) is,
therefore, mainly due to the effect of magnetism. \\
\indent In general, the presence of the stripe-type AFM order lifts the degeneracy of the phonon bands between 
[1 1 0] and \([1 \ \overline{1} \ 0]\) directions, which results in a small splitting of phonon modes.
In the present work, however, no phonon splitting, indicative of the AFM order, was observed.
This result is reasonable in the paramagnetic state, in which 
spin fluctuations can be assumed to be fast and the resulting phonon response becomes the average of that in the
magnetically-ordered state. 
Meanwhile if the magnetic order is stabilized, for example,
by pressure\cite{Bendele2010PhysRevLett,Bendele2012Phys.Rev.B,Kothapalli2016Nat.Commun,Bohmer2019PhysRevB},
the mode splitting is expected to emerge as in the case of the 122-type FeSCs\cite{Murai2016Phys.Rev.B}. \\
\indent Our results thus demonstrate that the inclusion of magnetism is crucial to describe the
structural properties of the paramagnetic phase in proximity to the magnetic instability. 
In relation to this, there has been increasing evidence that lattice dynamics properties of some materials, such as
elemental iron above the Curie temperature\cite{Kormann2012Phys.Rev.B} and iron silicide\cite{Krannich2015Nat.Commun.}, 
also exhibit a strong sensitivity to the fluctuating local moments existing in the paramagnetic state. 
In the present work, the effects of magnetism are treated at the static DFT level, and the inclusion of dynamical spin correlations is
highly desirable for a more realistic description of the structural properties.
An attempt in this direction has recently been made in Ref. [\onlinecite{Han2018Phys.Rev.Lett}]
by combining DFT with dynamical mean field theory (DFT+DMFT). \\
\indent The phase diagram of FeSe is quite distinct from that of all other known FeSCs
because its orthorhombic distortion is not accompanied by magnetic order\cite{Bohmer2017JPCM}.
The existence of the phase with decoupled lattice and spin degrees of freedom in FeSe 
has been interpreted as implying the importance of orbital ordering, particularly, in the context of nematicity. 
In contrast, our analysis reveals that the structural properties of FeSe are intimately coupled to the Fe-spin state
via fluctuating local moments in the paramagnetic state.
Such a dynamical aspect of magneto-elastic coupling has also been observed
in other FeSCs\cite{Fukuda2011Phhys.Rev.B,Reznik2009Phys.Rev.B,Murai2016Phys.Rev.B}. 
Hence, as far as the structural properties are concerned,
the physics of FeSe is more similar to that of other typical FeSCs than hitherto expected.\\
\indent We close the paper with some remarks on possible directions of future work.
In general, the presence of magnetism not only renormalizes the phonon frequencies, but also
enhances the electron-phonon (e-ph) coupling constant\cite{Boeri2010Phys.RevB., Coh2015New.J.Phys., Coh2016Phys.Rev.B}.
This enhancement is still not large enough to explain high-\(T_{\rm c}\) superconductivity in FeSCs, but is 
not negligibly small\cite{Boeri2010Phys.RevB.}.
In this regard, it is interesting to note that the e-ph properties of FeSe,
which match the pressure dependence of \(T_{\rm c}\),
can only be accounted for by including local spin fluctuations in the DFT+DMFT approach\cite{Mandal2014Phys.Rev.B},
indicating the possible interplay between magnetism and e-ph coupling. 
So far, the experimental verification of the enhanced e-ph coupling strength has been reported
for only one particular phonon mode\cite{Gerber2017Science}. 
The full momentum- and mode-resolved determination of the e-ph coupling demands considerable efforts combining
the experimental and computational methods.
This is an interesting topic for future IXS investigations on FeSCs.
\section{Summary}
To summarize, we performed a combined IXS and first-principles investigation of lattice dynamics properties in 
FeSe superconductor. We modeled the experimental phonon dispersion by imposing the AFM order within DFT
that can be attributed to the effect of fluctuating local moments.
Our analysis shows that, similar to typical FeSCs, the structural properties of FeSe are intimately coupled to the Fe-spin states.
Such a magneto-elastic coupling is a common feature that links together the various families of FeSCs. 
\section{Acknowledgments}
N.M. acknowledges valuable discussions with Prof. Koichi Kusakabe, Drs. Katsuhiro Suzuki, Kenji Nakajima, Ryoichi Kajimoto and Seiko Ohira-Kawamura.
This work was supported in part by Grants-in-Aid for Scientific Research (Grant No. JP19K14666), JSPS, Japan and
Grant for Basic Science Research Projects from the Sumitomo Foundation.
The synchrotron radiation experiments were performed at BL43LXU of SPring-8
with the approval of RIKEN.
All calculations were performed using the supercomputing system (SGI ICE X) of the Japan Atomic Energy Agency.
\bibliography{bibliography}

\begin{thebibliography}{48}%
\makeatletter
\providecommand \@ifxundefined [1]{%
 \@ifx{#1\undefined}
}%
\providecommand \@ifnum [1]{%
 \ifnum #1\expandafter \@firstoftwo
 \else \expandafter \@secondoftwo
 \fi
}%
\providecommand \@ifx [1]{%
 \ifx #1\expandafter \@firstoftwo
 \else \expandafter \@secondoftwo
 \fi
}%
\providecommand \natexlab [1]{#1}%
\providecommand \enquote  [1]{``#1''}%
\providecommand \bibnamefont  [1]{#1}%
\providecommand \bibfnamefont [1]{#1}%
\providecommand \citenamefont [1]{#1}%
\providecommand \href@noop [0]{\@secondoftwo}%
\providecommand \href [0]{\begingroup \@sanitize@url \@href}%
\providecommand \@href[1]{\@@startlink{#1}\@@href}%
\providecommand \@@href[1]{\endgroup#1\@@endlink}%
\providecommand \@sanitize@url [0]{\catcode `\\12\catcode `\$12\catcode
  `\&12\catcode `\#12\catcode `\^12\catcode `\_12\catcode `\%12\relax}%
\providecommand \@@startlink[1]{}%
\providecommand \@@endlink[0]{}%
\providecommand \url  [0]{\begingroup\@sanitize@url \@url }%
\providecommand \@url [1]{\endgroup\@href {#1}{\urlprefix }}%
\providecommand \urlprefix  [0]{URL }%
\providecommand \Eprint [0]{\href }%
\providecommand \doibase [0]{http://dx.doi.org/}%
\providecommand \selectlanguage [0]{\@gobble}%
\providecommand \bibinfo  [0]{\@secondoftwo}%
\providecommand \bibfield  [0]{\@secondoftwo}%
\providecommand \translation [1]{[#1]}%
\providecommand \BibitemOpen [0]{}%
\providecommand \bibitemStop [0]{}%
\providecommand \bibitemNoStop [0]{.\EOS\space}%
\providecommand \EOS [0]{\spacefactor3000\relax}%
\providecommand \BibitemShut  [1]{\csname bibitem#1\endcsname}%
\let\auto@bib@innerbib\@empty
\bibitem [{\citenamefont {Kamihara}\ \emph {et~al.}(2008)\citenamefont
  {Kamihara}, \citenamefont {Watanabe}, \citenamefont {Hirano},\ and\
  \citenamefont {Hosono}}]{Kamihara2008JACS}%
  \BibitemOpen
  \bibfield  {author} {\bibinfo {author} {\bibfnamefont {Y.}~\bibnamefont
  {Kamihara}}, \bibinfo {author} {\bibfnamefont {T.}~\bibnamefont {Watanabe}},
  \bibinfo {author} {\bibfnamefont {M.}~\bibnamefont {Hirano}}, \ and\ \bibinfo
  {author} {\bibfnamefont {H.}~\bibnamefont {Hosono}},\ }\href {\doibase 10.1021/ja800073m} {\bibfield  {journal} {\bibinfo  {journal} {J. Am. Chem.
  Soc.}\ }\textbf {\bibinfo {volume} {130}},\ \bibinfo {pages} {3296} (\bibinfo
  {year} {2008})}\BibitemShut {NoStop}%
\bibitem [{\citenamefont {Kreyssig}\ \emph {et~al.}(2008)\citenamefont
  {Kreyssig}, \citenamefont {Green}, \citenamefont {Lee}, \citenamefont
  {Samolyuk}, \citenamefont {Zajdel}, \citenamefont {Lynn}, \citenamefont
  {Bud'ko}, \citenamefont {Torikachvili}, \citenamefont {Ni}, \citenamefont
  {Nandi}, \citenamefont {Le\~ao}, \citenamefont {Poulton}, \citenamefont
  {Argyriou}, \citenamefont {Harmon}, \citenamefont {McQueeney}, \citenamefont
  {Canfield},\ and\ \citenamefont {Goldman}}]{Kreyssig2008Phys.Rev.B}%
  \BibitemOpen
  \bibfield  {author} {\bibinfo {author} {\bibfnamefont {A.}~\bibnamefont
  {Kreyssig}}, \bibinfo {author} {\bibfnamefont {M.~A.}\ \bibnamefont {Green}},
  \bibinfo {author} {\bibfnamefont {Y.}~\bibnamefont {Lee}}, \bibinfo {author}
  {\bibfnamefont {G.~D.}\ \bibnamefont {Samolyuk}}, \bibinfo {author}
  {\bibfnamefont {P.}~\bibnamefont {Zajdel}}, \bibinfo {author} {\bibfnamefont
  {J.~W.}\ \bibnamefont {Lynn}}, \bibinfo {author} {\bibfnamefont {S.~L.}\
  \bibnamefont {Bud'ko}}, \bibinfo {author} {\bibfnamefont {M.~S.}\
  \bibnamefont {Torikachvili}}, \bibinfo {author} {\bibfnamefont
  {N.}~\bibnamefont {Ni}}, \bibinfo {author} {\bibfnamefont {S.}~\bibnamefont
  {Nandi}}, \bibinfo {author} {\bibfnamefont {J.~B.}\ \bibnamefont {Le\~ao}},
  \bibinfo {author} {\bibfnamefont {S.~J.}\ \bibnamefont {Poulton}}, \bibinfo
  {author} {\bibfnamefont {D.~N.}\ \bibnamefont {Argyriou}}, \bibinfo {author}
  {\bibfnamefont {B.~N.}\ \bibnamefont {Harmon}}, \bibinfo {author}
  {\bibfnamefont {R.~J.}\ \bibnamefont {McQueeney}}, \bibinfo {author}
  {\bibfnamefont {P.~C.}\ \bibnamefont {Canfield}}, \ and\ \bibinfo {author}
  {\bibfnamefont {A.~I.}\ \bibnamefont {Goldman}},\ }\href {\doibase 10.1103/PhysRevB.78.184517} {\bibfield  {journal} {\bibinfo  {journal} {Phys.
  Rev. B}\ }\textbf {\bibinfo {volume} {78}},\ \bibinfo {pages} {184517}
  (\bibinfo {year} {2008})}\BibitemShut {NoStop}%
\bibitem [{\citenamefont {Fukuda}\ \emph {et~al.}(2008)\citenamefont {Fukuda},
  \citenamefont {Q.~R.~Baron}, \citenamefont {Shamoto}, \citenamefont
  {Ishikado}, \citenamefont {Nakamura}, \citenamefont {Machida}, \citenamefont
  {Uchiyama}, \citenamefont {Tsutsui}, \citenamefont {Iyo}, \citenamefont
  {Kito}, \citenamefont {Mizuki}, \citenamefont {Arai}, \citenamefont
  {Eisaki},\ and\ \citenamefont {Hosono}}]{Fukuda2008JPSJ}%
  \BibitemOpen
  \bibfield  {author} {\bibinfo {author} {\bibfnamefont {T.}~\bibnamefont
  {Fukuda}}, \bibinfo {author} {\bibfnamefont {A.}~\bibnamefont {Q.~R.~Baron}},
  \bibinfo {author} {\bibfnamefont {S.-i.}\ \bibnamefont {Shamoto}}, \bibinfo
  {author} {\bibfnamefont {M.}~\bibnamefont {Ishikado}}, \bibinfo {author}
  {\bibfnamefont {H.}~\bibnamefont {Nakamura}}, \bibinfo {author}
  {\bibfnamefont {M.}~\bibnamefont {Machida}}, \bibinfo {author} {\bibfnamefont
  {H.}~\bibnamefont {Uchiyama}}, \bibinfo {author} {\bibfnamefont
  {S.}~\bibnamefont {Tsutsui}}, \bibinfo {author} {\bibfnamefont
  {A.}~\bibnamefont {Iyo}}, \bibinfo {author} {\bibfnamefont {H.}~\bibnamefont
  {Kito}}, \bibinfo {author} {\bibfnamefont {J.}~\bibnamefont {Mizuki}},
  \bibinfo {author} {\bibfnamefont {M.}~\bibnamefont {Arai}}, \bibinfo {author}
  {\bibfnamefont {H.}~\bibnamefont {Eisaki}}, \ and\ \bibinfo {author}
  {\bibfnamefont {H.}~\bibnamefont {Hosono}},\ }\href {\doibase 10.1143/JPSJ.77.103715} {\bibfield  {journal} {\bibinfo  {journal} {J. Phys.
  Soc. Jpn.}\ }\textbf {\bibinfo {volume} {77}},\ \bibinfo {pages} {103715}
  (\bibinfo {year} {2008})}\BibitemShut {NoStop}%
\bibitem [{\citenamefont {Hahn}\ \emph {et~al.}(2009)\citenamefont {Hahn},
  \citenamefont {Lee}, \citenamefont {Ni}, \citenamefont {Canfield},
  \citenamefont {Goldman}, \citenamefont {McQueeney}, \citenamefont {Harmon},
  \citenamefont {Alatas}, \citenamefont {Leu}, \citenamefont {Alp},
  \citenamefont {Chung}, \citenamefont {Todorov},\ and\ \citenamefont
  {Kanatzidis}}]{Hahn2009PhysRevB}%
  \BibitemOpen
  \bibfield  {author} {\bibinfo {author} {\bibfnamefont {S.~E.}\ \bibnamefont
  {Hahn}}, \bibinfo {author} {\bibfnamefont {Y.}~\bibnamefont {Lee}}, \bibinfo
  {author} {\bibfnamefont {N.}~\bibnamefont {Ni}}, \bibinfo {author}
  {\bibfnamefont {P.~C.}\ \bibnamefont {Canfield}}, \bibinfo {author}
  {\bibfnamefont {A.~I.}\ \bibnamefont {Goldman}}, \bibinfo {author}
  {\bibfnamefont {R.~J.}\ \bibnamefont {McQueeney}}, \bibinfo {author}
  {\bibfnamefont {B.~N.}\ \bibnamefont {Harmon}}, \bibinfo {author}
  {\bibfnamefont {A.}~\bibnamefont {Alatas}}, \bibinfo {author} {\bibfnamefont
  {B.~M.}\ \bibnamefont {Leu}}, \bibinfo {author} {\bibfnamefont {E.~E.}\
  \bibnamefont {Alp}}, \bibinfo {author} {\bibfnamefont {D.~Y.}\ \bibnamefont
  {Chung}}, \bibinfo {author} {\bibfnamefont {I.~S.}\ \bibnamefont {Todorov}},
  \ and\ \bibinfo {author} {\bibfnamefont {M.~G.}\ \bibnamefont {Kanatzidis}},\
  }\href {\doibase 10.1103/PhysRevB.79.220511} {\bibfield  {journal} {\bibinfo
  {journal} {Phys. Rev. B}\ }\textbf {\bibinfo {volume} {79}},\ \bibinfo
  {pages} {220511} (\bibinfo {year} {2009})}\BibitemShut {NoStop}%
\bibitem [{\citenamefont {Reznik}\ \emph {et~al.}(2009)\citenamefont {Reznik},
  \citenamefont {Lokshin}, \citenamefont {Mitchell}, \citenamefont {Parshall},
  \citenamefont {Dmowski}, \citenamefont {Lamago}, \citenamefont {Heid},
  \citenamefont {Bohnen}, \citenamefont {Sefat}, \citenamefont {McGuire},
  \citenamefont {Sales}, \citenamefont {Mandrus}, \citenamefont {Subedi},
  \citenamefont {Singh}, \citenamefont {Alatas}, \citenamefont {Upton},
  \citenamefont {Said}, \citenamefont {Cunsolo}, \citenamefont {Shvyd'ko},\
  and\ \citenamefont {Egami}}]{Reznik2009Phys.Rev.B}%
  \BibitemOpen
  \bibfield  {author} {\bibinfo {author} {\bibfnamefont {D.}~\bibnamefont
  {Reznik}}, \bibinfo {author} {\bibfnamefont {K.}~\bibnamefont {Lokshin}},
  \bibinfo {author} {\bibfnamefont {D.~C.}\ \bibnamefont {Mitchell}}, \bibinfo
  {author} {\bibfnamefont {D.}~\bibnamefont {Parshall}}, \bibinfo {author}
  {\bibfnamefont {W.}~\bibnamefont {Dmowski}}, \bibinfo {author} {\bibfnamefont
  {D.}~\bibnamefont {Lamago}}, \bibinfo {author} {\bibfnamefont
  {R.}~\bibnamefont {Heid}}, \bibinfo {author} {\bibfnamefont {K.-P.}\
  \bibnamefont {Bohnen}}, \bibinfo {author} {\bibfnamefont {A.~S.}\
  \bibnamefont {Sefat}}, \bibinfo {author} {\bibfnamefont {M.~A.}\ \bibnamefont
  {McGuire}}, \bibinfo {author} {\bibfnamefont {B.~C.}\ \bibnamefont {Sales}},
  \bibinfo {author} {\bibfnamefont {D.~G.}\ \bibnamefont {Mandrus}}, \bibinfo
  {author} {\bibfnamefont {A.}~\bibnamefont {Subedi}}, \bibinfo {author}
  {\bibfnamefont {D.~J.}\ \bibnamefont {Singh}}, \bibinfo {author}
  {\bibfnamefont {A.}~\bibnamefont {Alatas}}, \bibinfo {author} {\bibfnamefont
  {M.~H.}\ \bibnamefont {Upton}}, \bibinfo {author} {\bibfnamefont {A.~H.}\
  \bibnamefont {Said}}, \bibinfo {author} {\bibfnamefont {A.}~\bibnamefont
  {Cunsolo}}, \bibinfo {author} {\bibfnamefont {Y.}~\bibnamefont {Shvyd'ko}}, \
  and\ \bibinfo {author} {\bibfnamefont {T.}~\bibnamefont {Egami}},\ }\href
  {\doibase 10.1103/PhysRevB.80.214534} {\bibfield  {journal} {\bibinfo
  {journal} {Phys. Rev. B}\ }\textbf {\bibinfo {volume} {80}},\ \bibinfo
  {pages} {214534} (\bibinfo {year} {2009})}\BibitemShut {NoStop}%
\bibitem [{\citenamefont {Fukuda}\ \emph {et~al.}(2011)\citenamefont {Fukuda},
  \citenamefont {Baron}, \citenamefont {Nakamura}, \citenamefont {Shamoto},
  \citenamefont {Ishikado}, \citenamefont {Machida}, \citenamefont {Uchiyama},
  \citenamefont {Iyo}, \citenamefont {Kito}, \citenamefont {Mizuki},
  \citenamefont {Arai},\ and\ \citenamefont {Eisaki}}]{Fukuda2011Phhys.Rev.B}%
  \BibitemOpen
  \bibfield  {author} {\bibinfo {author} {\bibfnamefont {T.}~\bibnamefont
  {Fukuda}}, \bibinfo {author} {\bibfnamefont {A.~Q.~R.}\ \bibnamefont
  {Baron}}, \bibinfo {author} {\bibfnamefont {H.}~\bibnamefont {Nakamura}},
  \bibinfo {author} {\bibfnamefont {S.}~\bibnamefont {Shamoto}}, \bibinfo
  {author} {\bibfnamefont {M.}~\bibnamefont {Ishikado}}, \bibinfo {author}
  {\bibfnamefont {M.}~\bibnamefont {Machida}}, \bibinfo {author} {\bibfnamefont
  {H.}~\bibnamefont {Uchiyama}}, \bibinfo {author} {\bibfnamefont
  {A.}~\bibnamefont {Iyo}}, \bibinfo {author} {\bibfnamefont {H.}~\bibnamefont
  {Kito}}, \bibinfo {author} {\bibfnamefont {J.}~\bibnamefont {Mizuki}},
  \bibinfo {author} {\bibfnamefont {M.}~\bibnamefont {Arai}}, \ and\ \bibinfo
  {author} {\bibfnamefont {H.}~\bibnamefont {Eisaki}},\ }\href {\doibase 10.1103/PhysRevB.84.064504} {\bibfield  {journal} {\bibinfo  {journal} {Phys.
  Rev. B}\ }\textbf {\bibinfo {volume} {84}},\ \bibinfo {pages} {064504}
  (\bibinfo {year} {2011})}\BibitemShut {NoStop}%
\bibitem [{\citenamefont {Murai}\ \emph {et~al.}(2016)\citenamefont {Murai},
  \citenamefont {Fukuda}, \citenamefont {Kobayashi}, \citenamefont {Nakajima},
  \citenamefont {Uchiyama}, \citenamefont {Ishikawa}, \citenamefont {Tsutsui},
  \citenamefont {Nakamura}, \citenamefont {Machida}, \citenamefont {Miyasaka},
  \citenamefont {Tajima},\ and\ \citenamefont {Baron}}]{Murai2016Phys.Rev.B}%
  \BibitemOpen
  \bibfield  {author} {\bibinfo {author} {\bibfnamefont {N.}~\bibnamefont
  {Murai}}, \bibinfo {author} {\bibfnamefont {T.}~\bibnamefont {Fukuda}},
  \bibinfo {author} {\bibfnamefont {T.}~\bibnamefont {Kobayashi}}, \bibinfo
  {author} {\bibfnamefont {M.}~\bibnamefont {Nakajima}}, \bibinfo {author}
  {\bibfnamefont {H.}~\bibnamefont {Uchiyama}}, \bibinfo {author}
  {\bibfnamefont {D.}~\bibnamefont {Ishikawa}}, \bibinfo {author}
  {\bibfnamefont {S.}~\bibnamefont {Tsutsui}}, \bibinfo {author} {\bibfnamefont
  {H.}~\bibnamefont {Nakamura}}, \bibinfo {author} {\bibfnamefont
  {M.}~\bibnamefont {Machida}}, \bibinfo {author} {\bibfnamefont
  {S.}~\bibnamefont {Miyasaka}}, \bibinfo {author} {\bibfnamefont
  {S.}~\bibnamefont {Tajima}}, \ and\ \bibinfo {author} {\bibfnamefont
  {A.~Q.~R.}\ \bibnamefont {Baron}},\ }\href {\doibase 10.1103/PhysRevB.93.020301} {\bibfield  {journal} {\bibinfo  {journal} {Phys.
  Rev. B}\ }\textbf {\bibinfo {volume} {93}},\ \bibinfo {pages} {020301}
  (\bibinfo {year} {2016})}\BibitemShut {NoStop}%
\bibitem [{\citenamefont {B\"ohmer}\ and\ \citenamefont
  {Kreisel}(2017)}]{Bohmer2017JPCM}%
  \BibitemOpen
  \bibfield  {author} {\bibinfo {author} {\bibfnamefont {A.~E.}\ \bibnamefont
  {B\"ohmer}}\ and\ \bibinfo {author} {\bibfnamefont {A.}~\bibnamefont
  {Kreisel}},\ }\href {\doibase 10.1088/1361-648x/aa9caa} {\bibfield  {journal}
  {\bibinfo  {journal} {J. Phys. Condens. Matter}\ }\textbf {\bibinfo {volume}
  {30}},\ \bibinfo {pages} {023001} (\bibinfo {year} {2017})}\BibitemShut
  {NoStop}%
\bibitem [{\citenamefont {B\"ohmer}\ \emph {et~al.}(2016)\citenamefont
  {B\"ohmer}, \citenamefont {Taufour}, \citenamefont {Straszheim},
  \citenamefont {Wolf},\ and\ \citenamefont
  {Canfield}}]{Bohmer2016PhysRevB.94.024526}%
  \BibitemOpen
  \bibfield  {author} {\bibinfo {author} {\bibfnamefont {A.~E.}\ \bibnamefont
  {B\"ohmer}}, \bibinfo {author} {\bibfnamefont {V.}~\bibnamefont {Taufour}},
  \bibinfo {author} {\bibfnamefont {W.~E.}\ \bibnamefont {Straszheim}},
  \bibinfo {author} {\bibfnamefont {T.}~\bibnamefont {Wolf}}, \ and\ \bibinfo
  {author} {\bibfnamefont {P.~C.}\ \bibnamefont {Canfield}},\ }\href {\doibase 10.1103/PhysRevB.94.024526} {\bibfield  {journal} {\bibinfo  {journal} {Phys.
  Rev. B}\ }\textbf {\bibinfo {volume} {94}},\ \bibinfo {pages} {024526}
  (\bibinfo {year} {2016})}\BibitemShut {NoStop}%
\bibitem [{\citenamefont {Q.~R.~Baron}(2010)}]{Baron2010riken}%
  \BibitemOpen
  \bibfield  {author} {\bibinfo {author} {\bibfnamefont {A.}~\bibnamefont
  {Q.~R.~Baron}},\ }\href
  {http://www.spring8.or.jp/pdf/ja/sp8-info/15-1-10/15-1-10-p14_press.pdf}
  {\bibfield  {journal} {\bibinfo  {journal} {SPring-8 Inf. Newsl}\ }\textbf
  {\bibinfo {volume} {15}},\ \bibinfo {pages} {14} (\bibinfo {year}
  {2010})}\BibitemShut {NoStop}%
\bibitem [{\citenamefont {Baron}(2016)}]{Baron2016}%
  \BibitemOpen
  \bibfield  {author} {\bibinfo {author} {\bibfnamefont {A.~Q.}\ \bibnamefont
  {Baron}},\ }\enquote {\bibinfo {title} {High-{R}esolution {I}nelastic
  {X}-{R}ay {S}cattering},}\ in\ \href@noop {} {\emph {\bibinfo {booktitle}
  {Synchrotron Light Sources and Free-Electron Lasers: Accelerator Physics,
  Instrumentation and Science Applications}}},\ \bibinfo {editor} {edited by\
  \bibinfo {editor} {\bibfnamefont {E.~J.}\ \bibnamefont {Jaeschke}}, \bibinfo
  {editor} {\bibfnamefont {S.}~\bibnamefont {Khan}}, \bibinfo {editor}
  {\bibfnamefont {J.~R.}\ \bibnamefont {Schneider}}, \ and\ \bibinfo {editor}
  {\bibfnamefont {J.~B.}\ \bibnamefont {Hastings}}}\ (\bibinfo  {publisher}
  {Springer International Publishing},\ \bibinfo {address} {Cham},\ \bibinfo
  {year} {2016})\ pp.\ \bibinfo {pages} {1643--1719}\BibitemShut {NoStop}%
\bibitem [{\citenamefont {Baron}\ \emph {et~al.}(2008)\citenamefont {Baron},
  \citenamefont {Sutter}, \citenamefont {Tsutsui}, \citenamefont {Uchiyama},
  \citenamefont {Masui}, \citenamefont {Tajima}, \citenamefont {Heid},\ and\
  \citenamefont {Bohnen}}]{Baron2008JPCM}%
  \BibitemOpen
  \bibfield  {author} {\bibinfo {author} {\bibfnamefont {A.~Q.~R.}\
  \bibnamefont {Baron}}, \bibinfo {author} {\bibfnamefont {J.~P.}\ \bibnamefont
  {Sutter}}, \bibinfo {author} {\bibfnamefont {S.}~\bibnamefont {Tsutsui}},
  \bibinfo {author} {\bibfnamefont {H.}~\bibnamefont {Uchiyama}}, \bibinfo
  {author} {\bibfnamefont {T.}~\bibnamefont {Masui}}, \bibinfo {author}
  {\bibfnamefont {S.}~\bibnamefont {Tajima}}, \bibinfo {author} {\bibfnamefont
  {R.}~\bibnamefont {Heid}}, \ and\ \bibinfo {author} {\bibfnamefont {K.-P.}\
  \bibnamefont {Bohnen}},\ }\href {\doibase 10.1016/j.jpcs.2008.06.119}
  {\bibfield  {journal} {\bibinfo  {journal} {J. Phys. Chem. Solids}\ }\textbf
  {\bibinfo {volume} {69}},\ \bibinfo {pages} {3100 } (\bibinfo {year}
  {2008})}\BibitemShut {NoStop}%
\bibitem [{\citenamefont {James}\ and\ \citenamefont
  {Roos}(1975)}]{James1975Comput.Phys.Commun.}%
  \BibitemOpen
  \bibfield  {author} {\bibinfo {author} {\bibfnamefont {F.}~\bibnamefont
  {James}}\ and\ \bibinfo {author} {\bibfnamefont {M.}~\bibnamefont {Roos}},\
  }\href {\doibase 10.1016/0010-4655(75)90039-9} {\bibfield  {journal}
  {\bibinfo  {journal} {Comput. Phys. Commun.}\ }\textbf {\bibinfo {volume}
  {10}},\ \bibinfo {pages} {343 } (\bibinfo {year} {1975})}\BibitemShut
  {NoStop}%
\bibitem [{\citenamefont {Momma}\ and\ \citenamefont
  {Izumi}(2011)}]{Momma2011}%
  \BibitemOpen
  \bibfield  {author} {\bibinfo {author} {\bibfnamefont {K.}~\bibnamefont
  {Momma}}\ and\ \bibinfo {author} {\bibfnamefont {F.}~\bibnamefont {Izumi}},\
  }\href {\doibase 10.1107/S0021889811038970} {\bibfield  {journal} {\bibinfo
  {journal} {J. Appl. Crystallogr.}\ }\textbf {\bibinfo {volume} {44}},\
  \bibinfo {pages} {1272} (\bibinfo {year} {2011})}\BibitemShut {NoStop}%
\bibitem [{\citenamefont {Baroni}\ \emph {et~al.}(2001)\citenamefont {Baroni},
  \citenamefont {de~Gironcoli}, \citenamefont {Dal~Corso},\ and\ \citenamefont
  {Giannozzi}}]{Baroni2001Rev.Mod.Phys}%
  \BibitemOpen
  \bibfield  {author} {\bibinfo {author} {\bibfnamefont {S.}~\bibnamefont
  {Baroni}}, \bibinfo {author} {\bibfnamefont {S.}~\bibnamefont
  {de~Gironcoli}}, \bibinfo {author} {\bibfnamefont {A.}~\bibnamefont
  {Dal~Corso}}, \ and\ \bibinfo {author} {\bibfnamefont {P.}~\bibnamefont
  {Giannozzi}},\ }\href {\doibase 10.1103/RevModPhys.73.515} {\bibfield
  {journal} {\bibinfo  {journal} {Rev. Mod. Phys.}\ }\textbf {\bibinfo {volume}
  {73}},\ \bibinfo {pages} {515} (\bibinfo {year} {2001})}\BibitemShut
  {NoStop}%
\bibitem [{\citenamefont {Giannozzi}\ \emph {et~al.}(2009)\citenamefont
  {Giannozzi}, \citenamefont {Baroni}, \citenamefont {Bonini}, \citenamefont
  {Calandra}, \citenamefont {Car}, \citenamefont {Cavazzoni}, \citenamefont
  {Ceresoli}, \citenamefont {Chiarotti}, \citenamefont {Cococcioni},
  \citenamefont {Dabo}, \citenamefont {Corso}, \citenamefont {de~Gironcoli},
  \citenamefont {Fabris}, \citenamefont {Fratesi}, \citenamefont {Gebauer},
  \citenamefont {Gerstmann}, \citenamefont {Gougoussis}, \citenamefont
  {Kokalj}, \citenamefont {Lazzeri}, \citenamefont {Martin-Samos},
  \citenamefont {Marzari}, \citenamefont {Mauri}, \citenamefont {Mazzarello},
  \citenamefont {Paolini}, \citenamefont {Pasquarello}, \citenamefont
  {Paulatto}, \citenamefont {Sbraccia}, \citenamefont {Scandolo}, \citenamefont
  {Sclauzero}, \citenamefont {Seitsonen}, \citenamefont {Smogunov},
  \citenamefont {Umari},\ and\ \citenamefont
  {Wentzcovitch}}]{Giannozzi2009J.Phys:Condensed.Matter}%
  \BibitemOpen
  \bibfield  {author} {\bibinfo {author} {\bibfnamefont {P.}~\bibnamefont
  {Giannozzi}}, \bibinfo {author} {\bibfnamefont {S.}~\bibnamefont {Baroni}},
  \bibinfo {author} {\bibfnamefont {N.}~\bibnamefont {Bonini}}, \bibinfo
  {author} {\bibfnamefont {M.}~\bibnamefont {Calandra}}, \bibinfo {author}
  {\bibfnamefont {R.}~\bibnamefont {Car}}, \bibinfo {author} {\bibfnamefont
  {C.}~\bibnamefont {Cavazzoni}}, \bibinfo {author} {\bibfnamefont
  {D.}~\bibnamefont {Ceresoli}}, \bibinfo {author} {\bibfnamefont {G.~L.}\
  \bibnamefont {Chiarotti}}, \bibinfo {author} {\bibfnamefont {M.}~\bibnamefont
  {Cococcioni}}, \bibinfo {author} {\bibfnamefont {I.}~\bibnamefont {Dabo}},
  \bibinfo {author} {\bibfnamefont {A.~D.}\ \bibnamefont {Corso}}, \bibinfo
  {author} {\bibfnamefont {S.}~\bibnamefont {de~Gironcoli}}, \bibinfo {author}
  {\bibfnamefont {S.}~\bibnamefont {Fabris}}, \bibinfo {author} {\bibfnamefont
  {G.}~\bibnamefont {Fratesi}}, \bibinfo {author} {\bibfnamefont
  {R.}~\bibnamefont {Gebauer}}, \bibinfo {author} {\bibfnamefont
  {U.}~\bibnamefont {Gerstmann}}, \bibinfo {author} {\bibfnamefont
  {C.}~\bibnamefont {Gougoussis}}, \bibinfo {author} {\bibfnamefont
  {A.}~\bibnamefont {Kokalj}}, \bibinfo {author} {\bibfnamefont
  {M.}~\bibnamefont {Lazzeri}}, \bibinfo {author} {\bibfnamefont
  {L.}~\bibnamefont {Martin-Samos}}, \bibinfo {author} {\bibfnamefont
  {N.}~\bibnamefont {Marzari}}, \bibinfo {author} {\bibfnamefont
  {F.}~\bibnamefont {Mauri}}, \bibinfo {author} {\bibfnamefont
  {R.}~\bibnamefont {Mazzarello}}, \bibinfo {author} {\bibfnamefont
  {S.}~\bibnamefont {Paolini}}, \bibinfo {author} {\bibfnamefont
  {A.}~\bibnamefont {Pasquarello}}, \bibinfo {author} {\bibfnamefont
  {L.}~\bibnamefont {Paulatto}}, \bibinfo {author} {\bibfnamefont
  {C.}~\bibnamefont {Sbraccia}}, \bibinfo {author} {\bibfnamefont
  {S.}~\bibnamefont {Scandolo}}, \bibinfo {author} {\bibfnamefont
  {G.}~\bibnamefont {Sclauzero}}, \bibinfo {author} {\bibfnamefont {A.~P.}\
  \bibnamefont {Seitsonen}}, \bibinfo {author} {\bibfnamefont {A.}~\bibnamefont
  {Smogunov}}, \bibinfo {author} {\bibfnamefont {P.}~\bibnamefont {Umari}}, \
  and\ \bibinfo {author} {\bibfnamefont {R.~M.}\ \bibnamefont {Wentzcovitch}},\
  }\href {\doibase 10.1088/0953-8984/21/39/395502} {\bibfield  {journal}
  {\bibinfo  {journal} {J. Phys. Condens. Matter}\ }\textbf {\bibinfo {volume}
  {21}},\ \bibinfo {pages} {395502} (\bibinfo {year} {2009})}\BibitemShut
  {NoStop}%
\bibitem [{\citenamefont {Giannozzi}\ \emph {et~al.}(2017)\citenamefont
  {Giannozzi}, \citenamefont {Andreussi}, \citenamefont {Brumme}, \citenamefont
  {Bunau}, \citenamefont {Nardelli}, \citenamefont {Calandra}, \citenamefont
  {Car}, \citenamefont {Cavazzoni}, \citenamefont {Ceresoli}, \citenamefont
  {Cococcioni}, \citenamefont {Colonna}, \citenamefont {Carnimeo},
  \citenamefont {Corso}, \citenamefont {de~Gironcoli}, \citenamefont {Delugas},
  \citenamefont {DiStasio}, \citenamefont {Ferretti}, \citenamefont {Floris},
  \citenamefont {Fratesi}, \citenamefont {Fugallo}, \citenamefont {Gebauer},
  \citenamefont {Gerstmann}, \citenamefont {Giustino}, \citenamefont {Gorni},
  \citenamefont {Jia}, \citenamefont {Kawamura}, \citenamefont {Ko},
  \citenamefont {Kokalj}, \citenamefont {K^^c3^^bc{\c{c}}^^c3^^bckbenli},
  \citenamefont {Lazzeri}, \citenamefont {Marsili}, \citenamefont {Marzari},
  \citenamefont {Mauri}, \citenamefont {Nguyen}, \citenamefont {Nguyen},
  \citenamefont {de-la Roza}, \citenamefont {Paulatto}, \citenamefont
  {Ponc{\'{e}}}, \citenamefont {Rocca}, \citenamefont {Sabatini}, \citenamefont
  {Santra}, \citenamefont {Schlipf}, \citenamefont {Seitsonen}, \citenamefont
  {Smogunov}, \citenamefont {Timrov}, \citenamefont {Thonhauser}, \citenamefont
  {Umari}, \citenamefont {Vast}, \citenamefont {Wu},\ and\ \citenamefont
  {Baroni}}]{Giannozzi2017J.Phys:Condensed.Matter}%
  \BibitemOpen
  \bibfield  {author} {\bibinfo {author} {\bibfnamefont {P.}~\bibnamefont
  {Giannozzi}}, \bibinfo {author} {\bibfnamefont {O.}~\bibnamefont
  {Andreussi}}, \bibinfo {author} {\bibfnamefont {T.}~\bibnamefont {Brumme}},
  \bibinfo {author} {\bibfnamefont {O.}~\bibnamefont {Bunau}}, \bibinfo
  {author} {\bibfnamefont {M.~B.}\ \bibnamefont {Nardelli}}, \bibinfo {author}
  {\bibfnamefont {M.}~\bibnamefont {Calandra}}, \bibinfo {author}
  {\bibfnamefont {R.}~\bibnamefont {Car}}, \bibinfo {author} {\bibfnamefont
  {C.}~\bibnamefont {Cavazzoni}}, \bibinfo {author} {\bibfnamefont
  {D.}~\bibnamefont {Ceresoli}}, \bibinfo {author} {\bibfnamefont
  {M.}~\bibnamefont {Cococcioni}}, \bibinfo {author} {\bibfnamefont
  {N.}~\bibnamefont {Colonna}}, \bibinfo {author} {\bibfnamefont
  {I.}~\bibnamefont {Carnimeo}}, \bibinfo {author} {\bibfnamefont {A.~D.}\
  \bibnamefont {Corso}}, \bibinfo {author} {\bibfnamefont {S.}~\bibnamefont
  {de~Gironcoli}}, \bibinfo {author} {\bibfnamefont {P.}~\bibnamefont
  {Delugas}}, \bibinfo {author} {\bibfnamefont {R.~A.}\ \bibnamefont
  {DiStasio}}, \bibinfo {author} {\bibfnamefont {A.}~\bibnamefont {Ferretti}},
  \bibinfo {author} {\bibfnamefont {A.}~\bibnamefont {Floris}}, \bibinfo
  {author} {\bibfnamefont {G.}~\bibnamefont {Fratesi}}, \bibinfo {author}
  {\bibfnamefont {G.}~\bibnamefont {Fugallo}}, \bibinfo {author} {\bibfnamefont
  {R.}~\bibnamefont {Gebauer}}, \bibinfo {author} {\bibfnamefont
  {U.}~\bibnamefont {Gerstmann}}, \bibinfo {author} {\bibfnamefont
  {F.}~\bibnamefont {Giustino}}, \bibinfo {author} {\bibfnamefont
  {T.}~\bibnamefont {Gorni}}, \bibinfo {author} {\bibfnamefont
  {J.}~\bibnamefont {Jia}}, \bibinfo {author} {\bibfnamefont {M.}~\bibnamefont
  {Kawamura}}, \bibinfo {author} {\bibfnamefont {H.-Y.}\ \bibnamefont {Ko}},
  \bibinfo {author} {\bibfnamefont {A.}~\bibnamefont {Kokalj}}, \bibinfo
  {author} {\bibfnamefont {E.}~\bibnamefont {K^^c3^^bc{\c{c}}^^c3^^bckbenli}},
  \bibinfo {author} {\bibfnamefont {M.}~\bibnamefont {Lazzeri}}, \bibinfo
  {author} {\bibfnamefont {M.}~\bibnamefont {Marsili}}, \bibinfo {author}
  {\bibfnamefont {N.}~\bibnamefont {Marzari}}, \bibinfo {author} {\bibfnamefont
  {F.}~\bibnamefont {Mauri}}, \bibinfo {author} {\bibfnamefont {N.~L.}\
  \bibnamefont {Nguyen}}, \bibinfo {author} {\bibfnamefont {H.-V.}\
  \bibnamefont {Nguyen}}, \bibinfo {author} {\bibfnamefont {A.~O.}\
  \bibnamefont {de-la Roza}}, \bibinfo {author} {\bibfnamefont
  {L.}~\bibnamefont {Paulatto}}, \bibinfo {author} {\bibfnamefont
  {S.}~\bibnamefont {Ponc{\'{e}}}}, \bibinfo {author} {\bibfnamefont
  {D.}~\bibnamefont {Rocca}}, \bibinfo {author} {\bibfnamefont
  {R.}~\bibnamefont {Sabatini}}, \bibinfo {author} {\bibfnamefont
  {B.}~\bibnamefont {Santra}}, \bibinfo {author} {\bibfnamefont
  {M.}~\bibnamefont {Schlipf}}, \bibinfo {author} {\bibfnamefont {A.~P.}\
  \bibnamefont {Seitsonen}}, \bibinfo {author} {\bibfnamefont {A.}~\bibnamefont
  {Smogunov}}, \bibinfo {author} {\bibfnamefont {I.}~\bibnamefont {Timrov}},
  \bibinfo {author} {\bibfnamefont {T.}~\bibnamefont {Thonhauser}}, \bibinfo
  {author} {\bibfnamefont {P.}~\bibnamefont {Umari}}, \bibinfo {author}
  {\bibfnamefont {N.}~\bibnamefont {Vast}}, \bibinfo {author} {\bibfnamefont
  {X.}~\bibnamefont {Wu}}, \ and\ \bibinfo {author} {\bibfnamefont
  {S.}~\bibnamefont {Baroni}},\ }\href {\doibase 10.1088/1361-648x/aa8f79}
  {\bibfield  {journal} {\bibinfo  {journal} {J. Phys. Condens. Matter}\
  }\textbf {\bibinfo {volume} {29}},\ \bibinfo {pages} {465901} (\bibinfo
  {year} {2017})}\BibitemShut {NoStop}%
\bibitem [{\citenamefont {Perdew}\ \emph {et~al.}(2008)\citenamefont {Perdew},
  \citenamefont {Ruzsinszky}, \citenamefont {Csonka}, \citenamefont {Vydrov},
  \citenamefont {Scuseria}, \citenamefont {Constantin}, \citenamefont {Zhou},\
  and\ \citenamefont {Burke}}]{Perdew2008Phys.Rev.Lett}%
  \BibitemOpen
  \bibfield  {author} {\bibinfo {author} {\bibfnamefont {J.~P.}\ \bibnamefont
  {Perdew}}, \bibinfo {author} {\bibfnamefont {A.}~\bibnamefont {Ruzsinszky}},
  \bibinfo {author} {\bibfnamefont {G.~I.}\ \bibnamefont {Csonka}}, \bibinfo
  {author} {\bibfnamefont {O.~A.}\ \bibnamefont {Vydrov}}, \bibinfo {author}
  {\bibfnamefont {G.~E.}\ \bibnamefont {Scuseria}}, \bibinfo {author}
  {\bibfnamefont {L.~A.}\ \bibnamefont {Constantin}}, \bibinfo {author}
  {\bibfnamefont {X.}~\bibnamefont {Zhou}}, \ and\ \bibinfo {author}
  {\bibfnamefont {K.}~\bibnamefont {Burke}},\ }\href {\doibase 10.1103/PhysRevLett.100.136406} {\bibfield  {journal} {\bibinfo  {journal}
  {Phys. Rev. Lett.}\ }\textbf {\bibinfo {volume} {100}},\ \bibinfo {pages}
  {136406} (\bibinfo {year} {2008})}\BibitemShut {NoStop}%
\bibitem [{\citenamefont {Corso}(2014)}]{DalCorso2014Comput.Mater.Sci.}%
  \BibitemOpen
  \bibfield  {author} {\bibinfo {author} {\bibfnamefont {A.~D.}\ \bibnamefont
  {Corso}},\ }\href {\doibase 10.1016/j.commatsci.2014.07.043} {\bibfield
  {journal} {\bibinfo  {journal} {Comput. Mater. Sci.}\ }\textbf {\bibinfo
  {volume} {95}},\ \bibinfo {pages} {337 } (\bibinfo {year}
  {2014})}\BibitemShut {NoStop}%
\bibitem [{\citenamefont {Matsuura}\ \emph {et~al.}(2017)\citenamefont
  {Matsuura}, \citenamefont {Mizukami}, \citenamefont {Arai}, \citenamefont
  {Sugimura}, \citenamefont {Maejima}, \citenamefont {Machida}, \citenamefont
  {Watanuki}, \citenamefont {Fukuda}, \citenamefont {Yajima}, \citenamefont
  {Hiroi}, \citenamefont {Yip}, \citenamefont {Chan}, \citenamefont {Niu},
  \citenamefont {Hosoi}, \citenamefont {Ishida}, \citenamefont {Mukasa},
  \citenamefont {Kasahara}, \citenamefont {Cheng}, \citenamefont {Goh},
  \citenamefont {Matsuda}, \citenamefont {Uwatoko},\ and\ \citenamefont
  {Shibauchi}}]{Matsuura2017Nat.Commun}%
  \BibitemOpen
  \bibfield  {author} {\bibinfo {author} {\bibfnamefont {K.}~\bibnamefont
  {Matsuura}}, \bibinfo {author} {\bibfnamefont {Y.}~\bibnamefont {Mizukami}},
  \bibinfo {author} {\bibfnamefont {Y.}~\bibnamefont {Arai}}, \bibinfo {author}
  {\bibfnamefont {Y.}~\bibnamefont {Sugimura}}, \bibinfo {author}
  {\bibfnamefont {N.}~\bibnamefont {Maejima}}, \bibinfo {author} {\bibfnamefont
  {A.}~\bibnamefont {Machida}}, \bibinfo {author} {\bibfnamefont
  {T.}~\bibnamefont {Watanuki}}, \bibinfo {author} {\bibfnamefont
  {T.}~\bibnamefont {Fukuda}}, \bibinfo {author} {\bibfnamefont
  {T.}~\bibnamefont {Yajima}}, \bibinfo {author} {\bibfnamefont
  {Z.}~\bibnamefont {Hiroi}}, \bibinfo {author} {\bibfnamefont {K.~Y.}\
  \bibnamefont {Yip}}, \bibinfo {author} {\bibfnamefont {Y.~C.}\ \bibnamefont
  {Chan}}, \bibinfo {author} {\bibfnamefont {Q.}~\bibnamefont {Niu}}, \bibinfo
  {author} {\bibfnamefont {S.}~\bibnamefont {Hosoi}}, \bibinfo {author}
  {\bibfnamefont {K.}~\bibnamefont {Ishida}}, \bibinfo {author} {\bibfnamefont
  {K.}~\bibnamefont {Mukasa}}, \bibinfo {author} {\bibfnamefont
  {S.}~\bibnamefont {Kasahara}}, \bibinfo {author} {\bibfnamefont {J.~G.}\
  \bibnamefont {Cheng}}, \bibinfo {author} {\bibfnamefont {S.~K.}\ \bibnamefont
  {Goh}}, \bibinfo {author} {\bibfnamefont {Y.}~\bibnamefont {Matsuda}},
  \bibinfo {author} {\bibfnamefont {Y.}~\bibnamefont {Uwatoko}}, \ and\
  \bibinfo {author} {\bibfnamefont {T.}~\bibnamefont {Shibauchi}},\ }\href
  {\doibase 10.1038/s41467-017-01277-x} {\bibfield  {journal} {\bibinfo
  {journal} {Nat. Commun.}\ }\textbf {\bibinfo {volume} {8}},\ \bibinfo {pages}
  {1143} (\bibinfo {year} {2017})}\BibitemShut {NoStop}%
\bibitem [{Note1()}]{Note1}%
  \BibitemOpen
  \bibinfo {note} {Due to the finite momentum acceptance of the analyzers, some
  unexpected modes appear in the measured spectra (e.g, transverse modes in the
  longitudinal directions). To facilitate an easy comparison between theory and
  experiment, we here plot the experimental data expected in the measured
  symmetry directions.}\BibitemShut {Stop}%
\bibitem [{\citenamefont {Squires}(2012)}]{squires_2012}%
  \BibitemOpen
  \bibfield  {author} {\bibinfo {author} {\bibfnamefont {G.~L.}\ \bibnamefont
  {Squires}},\ }\href@noop {} {\emph {\bibinfo {title} {Introduction to the
  Theory of Thermal Neutron Scattering}}},\ \bibinfo {edition} {3rd}\ ed.\
  (\bibinfo  {publisher} {Cambridge University Press},\ \bibinfo {year}
  {2012})\BibitemShut {NoStop}%
\bibitem [{\citenamefont {Watson}\ \emph {et~al.}(2019)\citenamefont {Watson},
  \citenamefont {Dudin}, \citenamefont {Rhodes}, \citenamefont {Evtushinsky},
  \citenamefont {Iwasawa}, \citenamefont {Aswartham}, \citenamefont {Wurmehl},
  \citenamefont {B{\"u}chner}, \citenamefont {Hoesch},\ and\ \citenamefont
  {Kim}}]{Watson2019Quantum_Materials}%
  \BibitemOpen
  \bibfield  {author} {\bibinfo {author} {\bibfnamefont {M.~D.}\ \bibnamefont
  {Watson}}, \bibinfo {author} {\bibfnamefont {P.}~\bibnamefont {Dudin}},
  \bibinfo {author} {\bibfnamefont {L.~C.}\ \bibnamefont {Rhodes}}, \bibinfo
  {author} {\bibfnamefont {D.~V.}\ \bibnamefont {Evtushinsky}}, \bibinfo
  {author} {\bibfnamefont {H.}~\bibnamefont {Iwasawa}}, \bibinfo {author}
  {\bibfnamefont {S.}~\bibnamefont {Aswartham}}, \bibinfo {author}
  {\bibfnamefont {S.}~\bibnamefont {Wurmehl}}, \bibinfo {author} {\bibfnamefont
  {B.}~\bibnamefont {B{\"u}chner}}, \bibinfo {author} {\bibfnamefont
  {M.}~\bibnamefont {Hoesch}}, \ and\ \bibinfo {author} {\bibfnamefont {T.~K.}\
  \bibnamefont {Kim}},\ }\href {\doibase 10.1038/s41535-019-0174-z} {\bibfield
  {journal} {\bibinfo  {journal} {npj Quantum Materials}\ }\textbf {\bibinfo
  {volume} {4}},\ \bibinfo {pages} {36} (\bibinfo {year} {2019})}\BibitemShut
  {NoStop}%
\bibitem [{\citenamefont {Fedorov}\ \emph {et~al.}(2019)\citenamefont
  {Fedorov}, \citenamefont {Yaresko}, \citenamefont {Haubold}, \citenamefont
  {Kushnirenko}, \citenamefont {Kim}, \citenamefont {B\"uchner}, \citenamefont
  {Aswartham}, \citenamefont {Wurmehl},\ and\ \citenamefont
  {Borisenko}}]{Fedorov2019PhysRevB}%
  \BibitemOpen
  \bibfield  {author} {\bibinfo {author} {\bibfnamefont {A.}~\bibnamefont
  {Fedorov}}, \bibinfo {author} {\bibfnamefont {A.}~\bibnamefont {Yaresko}},
  \bibinfo {author} {\bibfnamefont {E.}~\bibnamefont {Haubold}}, \bibinfo
  {author} {\bibfnamefont {Y.}~\bibnamefont {Kushnirenko}}, \bibinfo {author}
  {\bibfnamefont {T.}~\bibnamefont {Kim}}, \bibinfo {author} {\bibfnamefont
  {B.}~\bibnamefont {B\"uchner}}, \bibinfo {author} {\bibfnamefont
  {S.}~\bibnamefont {Aswartham}}, \bibinfo {author} {\bibfnamefont
  {S.}~\bibnamefont {Wurmehl}}, \ and\ \bibinfo {author} {\bibfnamefont
  {S.}~\bibnamefont {Borisenko}},\ }\href {\doibase 10.1103/PhysRevB.100.024517} {\bibfield  {journal} {\bibinfo  {journal}
  {Phys. Rev. B}\ }\textbf {\bibinfo {volume} {100}},\ \bibinfo {pages}
  {024517} (\bibinfo {year} {2019})}\BibitemShut {NoStop}%
\bibitem [{\citenamefont {Ptok}\ \emph {et~al.}(2019)\citenamefont {Ptok},
  \citenamefont {Sternik}, \citenamefont {Kapcia},\ and\ \citenamefont
  {Piekarz}}]{2019PtoKPhysRevB}%
  \BibitemOpen
  \bibfield  {author} {\bibinfo {author} {\bibfnamefont {A.}~\bibnamefont
  {Ptok}}, \bibinfo {author} {\bibfnamefont {M.}~\bibnamefont {Sternik}},
  \bibinfo {author} {\bibfnamefont {K.~J.}\ \bibnamefont {Kapcia}}, \ and\
  \bibinfo {author} {\bibfnamefont {P.}~\bibnamefont {Piekarz}},\ }\href
  {\doibase 10.1103/PhysRevB.99.134103} {\bibfield  {journal} {\bibinfo
  {journal} {Phys. Rev. B}\ }\textbf {\bibinfo {volume} {99}},\ \bibinfo
  {pages} {134103} (\bibinfo {year} {2019})}\BibitemShut {NoStop}%
\bibitem [{\citenamefont {Yildirim}(2008)}]{Yildirim2008PhysRevLett}%
  \BibitemOpen
  \bibfield  {author} {\bibinfo {author} {\bibfnamefont {T.}~\bibnamefont
  {Yildirim}},\ }\href {\doibase 10.1103/PhysRevLett.101.057010} {\bibfield
  {journal} {\bibinfo  {journal} {Phys. Rev. Lett.}\ }\textbf {\bibinfo
  {volume} {101}},\ \bibinfo {pages} {057010} (\bibinfo {year}
  {2008})}\BibitemShut {NoStop}%
\bibitem [{\citenamefont {Ishibashi}\ \emph {et~al.}(2008)\citenamefont
  {Ishibashi}, \citenamefont {Terakura},\ and\ \citenamefont
  {Hosono}}]{Ishibashi2008JPSJ}%
  \BibitemOpen
  \bibfield  {author} {\bibinfo {author} {\bibfnamefont {S.}~\bibnamefont
  {Ishibashi}}, \bibinfo {author} {\bibfnamefont {K.}~\bibnamefont {Terakura}},
  \ and\ \bibinfo {author} {\bibfnamefont {H.}~\bibnamefont {Hosono}},\ }\href
  {\doibase 10.1143/JPSJ.77.053709} {\bibfield  {journal} {\bibinfo  {journal}
  {J. Phys. Soc. Jpn.}\ }\textbf {\bibinfo {volume} {77}},\ \bibinfo {pages}
  {053709} (\bibinfo {year} {2008})}\BibitemShut {NoStop}%
\bibitem [{\citenamefont {Mazin}\ and\ \citenamefont
  {Johannes}(2008)}]{Mazin2008Nat.Phys.}%
  \BibitemOpen
  \bibfield  {author} {\bibinfo {author} {\bibfnamefont {I.~I.}\ \bibnamefont
  {Mazin}}\ and\ \bibinfo {author} {\bibfnamefont {M.~D.}\ \bibnamefont
  {Johannes}},\ }\href {https://doi.org/10.1038/nphys1160} {\bibfield
  {journal} {\bibinfo  {journal} {Nat. Phys.}\ }\textbf {\bibinfo {volume}
  {5}},\ \bibinfo {pages} {141} (\bibinfo {year} {2008})}\BibitemShut {NoStop}%
\bibitem [{\citenamefont {Abrikosov}\ \emph {et~al.}(2016)\citenamefont
  {Abrikosov}, \citenamefont {Ponomareva}, \citenamefont {Steneteg},
  \citenamefont {Barannikova},\ and\ \citenamefont {Alling}}]{Abrikosov2016}%
  \BibitemOpen
  \bibfield  {author} {\bibinfo {author} {\bibfnamefont {I.}~\bibnamefont
  {Abrikosov}}, \bibinfo {author} {\bibfnamefont {A.}~\bibnamefont
  {Ponomareva}}, \bibinfo {author} {\bibfnamefont {P.}~\bibnamefont
  {Steneteg}}, \bibinfo {author} {\bibfnamefont {S.}~\bibnamefont
  {Barannikova}}, \ and\ \bibinfo {author} {\bibfnamefont {B.}~\bibnamefont
  {Alling}},\ }\href {\doibase 10.1016/j.cossms.2015.07.003} {\bibfield
  {journal} {\bibinfo  {journal} {Curr. Opin. Solid State Mater. Sci.}\
  }\textbf {\bibinfo {volume} {20}},\ \bibinfo {pages} {85 } (\bibinfo {year}
  {2016})}\BibitemShut {NoStop}%
\bibitem [{\citenamefont {Rahn}\ \emph {et~al.}(2015)\citenamefont {Rahn},
  \citenamefont {Ewings}, \citenamefont {Sedlmaier}, \citenamefont {Clarke},\
  and\ \citenamefont {Boothroyd}}]{Boothroyd2015Phys.RevB}%
  \BibitemOpen
  \bibfield  {author} {\bibinfo {author} {\bibfnamefont {M.~C.}\ \bibnamefont
  {Rahn}}, \bibinfo {author} {\bibfnamefont {R.~A.}\ \bibnamefont {Ewings}},
  \bibinfo {author} {\bibfnamefont {S.~J.}\ \bibnamefont {Sedlmaier}}, \bibinfo
  {author} {\bibfnamefont {S.~J.}\ \bibnamefont {Clarke}}, \ and\ \bibinfo
  {author} {\bibfnamefont {A.~T.}\ \bibnamefont {Boothroyd}},\ }\href {\doibase 10.1103/PhysRevB.91.180501} {\bibfield  {journal} {\bibinfo  {journal} {Phys.
  Rev. B}\ }\textbf {\bibinfo {volume} {91}},\ \bibinfo {pages} {180501}
  (\bibinfo {year} {2015})}\BibitemShut {NoStop}%
\bibitem [{\citenamefont {Wang}\ \emph {et~al.}(2015)\citenamefont {Wang},
  \citenamefont {Shen}, \citenamefont {Pan}, \citenamefont {Hao}, \citenamefont
  {Ma}, \citenamefont {Zhou}, \citenamefont {Steffens}, \citenamefont
  {Schmalzl}, \citenamefont {Forrest}, \citenamefont {Abdel-Hafiez},
  \citenamefont {Chen}, \citenamefont {Chareev}, \citenamefont {Vasiliev},
  \citenamefont {Bourges}, \citenamefont {Sidis}, \citenamefont {Cao},\ and\
  \citenamefont {Zhao}}]{Wang2015Nat.Mater}%
  \BibitemOpen
  \bibfield  {author} {\bibinfo {author} {\bibfnamefont {Q.}~\bibnamefont
  {Wang}}, \bibinfo {author} {\bibfnamefont {Y.}~\bibnamefont {Shen}}, \bibinfo
  {author} {\bibfnamefont {B.}~\bibnamefont {Pan}}, \bibinfo {author}
  {\bibfnamefont {Y.}~\bibnamefont {Hao}}, \bibinfo {author} {\bibfnamefont
  {M.}~\bibnamefont {Ma}}, \bibinfo {author} {\bibfnamefont {F.}~\bibnamefont
  {Zhou}}, \bibinfo {author} {\bibfnamefont {P.}~\bibnamefont {Steffens}},
  \bibinfo {author} {\bibfnamefont {K.}~\bibnamefont {Schmalzl}}, \bibinfo
  {author} {\bibfnamefont {T.~R.}\ \bibnamefont {Forrest}}, \bibinfo {author}
  {\bibfnamefont {M.}~\bibnamefont {Abdel-Hafiez}}, \bibinfo {author}
  {\bibfnamefont {X.}~\bibnamefont {Chen}}, \bibinfo {author} {\bibfnamefont
  {D.~A.}\ \bibnamefont {Chareev}}, \bibinfo {author} {\bibfnamefont {A.~N.}\
  \bibnamefont {Vasiliev}}, \bibinfo {author} {\bibfnamefont {P.}~\bibnamefont
  {Bourges}}, \bibinfo {author} {\bibfnamefont {Y.}~\bibnamefont {Sidis}},
  \bibinfo {author} {\bibfnamefont {H.}~\bibnamefont {Cao}}, \ and\ \bibinfo
  {author} {\bibfnamefont {J.}~\bibnamefont {Zhao}},\ }\href
  {https://doi.org/10.1038/nmat4492} {\bibfield  {journal} {\bibinfo  {journal}
  {Nat. Mater.}\ }\textbf {\bibinfo {volume} {15}},\ \bibinfo {pages} {159}
  (\bibinfo {year} {2015})}\BibitemShut {NoStop}%
\bibitem [{\citenamefont {Wang}\ \emph {et~al.}(2016)\citenamefont {Wang},
  \citenamefont {Shen}, \citenamefont {Pan}, \citenamefont {Zhang},
  \citenamefont {Ikeuchi}, \citenamefont {Iida}, \citenamefont {Christianson},
  \citenamefont {Walker}, \citenamefont {Adroja}, \citenamefont {Abdel-Hafiez},
  \citenamefont {Chen}, \citenamefont {Chareev}, \citenamefont {Vasiliev},\
  and\ \citenamefont {Zhao}}]{Wang2016Nat.Commun.}%
  \BibitemOpen
  \bibfield  {author} {\bibinfo {author} {\bibfnamefont {Q.}~\bibnamefont
  {Wang}}, \bibinfo {author} {\bibfnamefont {Y.}~\bibnamefont {Shen}}, \bibinfo
  {author} {\bibfnamefont {B.}~\bibnamefont {Pan}}, \bibinfo {author}
  {\bibfnamefont {X.}~\bibnamefont {Zhang}}, \bibinfo {author} {\bibfnamefont
  {K.}~\bibnamefont {Ikeuchi}}, \bibinfo {author} {\bibfnamefont
  {K.}~\bibnamefont {Iida}}, \bibinfo {author} {\bibfnamefont {A.~D.}\
  \bibnamefont {Christianson}}, \bibinfo {author} {\bibfnamefont {H.~C.}\
  \bibnamefont {Walker}}, \bibinfo {author} {\bibfnamefont {D.~T.}\
  \bibnamefont {Adroja}}, \bibinfo {author} {\bibfnamefont {M.}~\bibnamefont
  {Abdel-Hafiez}}, \bibinfo {author} {\bibfnamefont {X.}~\bibnamefont {Chen}},
  \bibinfo {author} {\bibfnamefont {D.~A.}\ \bibnamefont {Chareev}}, \bibinfo
  {author} {\bibfnamefont {A.~N.}\ \bibnamefont {Vasiliev}}, \ and\ \bibinfo
  {author} {\bibfnamefont {J.}~\bibnamefont {Zhao}},\ }\href
  {https://doi.org/10.1038/ncomms12182} {\bibfield  {journal} {\bibinfo
  {journal} {Nat. Commun.}\ }\textbf {\bibinfo {volume} {7}},\ \bibinfo {pages}
  {12182} (\bibinfo {year} {2016})}\BibitemShut {NoStop}%
\bibitem [{\citenamefont {Ma}\ \emph {et~al.}(2017)\citenamefont {Ma},
  \citenamefont {Bourges}, \citenamefont {Sidis}, \citenamefont {Xu},
  \citenamefont {Li}, \citenamefont {Hu}, \citenamefont {Li}, \citenamefont
  {Wang},\ and\ \citenamefont {Li}}]{Ma2017Phys.Rev.X}%
  \BibitemOpen
  \bibfield  {author} {\bibinfo {author} {\bibfnamefont {M.}~\bibnamefont
  {Ma}}, \bibinfo {author} {\bibfnamefont {P.}~\bibnamefont {Bourges}},
  \bibinfo {author} {\bibfnamefont {Y.}~\bibnamefont {Sidis}}, \bibinfo
  {author} {\bibfnamefont {Y.}~\bibnamefont {Xu}}, \bibinfo {author}
  {\bibfnamefont {S.}~\bibnamefont {Li}}, \bibinfo {author} {\bibfnamefont
  {B.}~\bibnamefont {Hu}}, \bibinfo {author} {\bibfnamefont {J.}~\bibnamefont
  {Li}}, \bibinfo {author} {\bibfnamefont {F.}~\bibnamefont {Wang}}, \ and\
  \bibinfo {author} {\bibfnamefont {Y.}~\bibnamefont {Li}},\ }\href {\doibase 10.1103/PhysRevX.7.021025} {\bibfield  {journal} {\bibinfo  {journal} {Phys.
  Rev. X}\ }\textbf {\bibinfo {volume} {7}},\ \bibinfo {pages} {021025}
  (\bibinfo {year} {2017})}\BibitemShut {NoStop}%
\bibitem [{\citenamefont {Chen}\ \emph {et~al.}(2019)\citenamefont {Chen},
  \citenamefont {Chen}, \citenamefont {Kreisel}, \citenamefont {Lu},
  \citenamefont {Schneidewind}, \citenamefont {Qiu}, \citenamefont {Park},
  \citenamefont {Perring}, \citenamefont {Stewart}, \citenamefont {Cao},
  \citenamefont {Zhang}, \citenamefont {Li}, \citenamefont {Rong},
  \citenamefont {Wei}, \citenamefont {Andersen}, \citenamefont {Hirschfeld},
  \citenamefont {Broholm},\ and\ \citenamefont {Dai}}]{Chen2019Nat.Mater}%
  \BibitemOpen
  \bibfield  {author} {\bibinfo {author} {\bibfnamefont {T.}~\bibnamefont
  {Chen}}, \bibinfo {author} {\bibfnamefont {Y.}~\bibnamefont {Chen}}, \bibinfo
  {author} {\bibfnamefont {A.}~\bibnamefont {Kreisel}}, \bibinfo {author}
  {\bibfnamefont {X.}~\bibnamefont {Lu}}, \bibinfo {author} {\bibfnamefont
  {A.}~\bibnamefont {Schneidewind}}, \bibinfo {author} {\bibfnamefont
  {Y.}~\bibnamefont {Qiu}}, \bibinfo {author} {\bibfnamefont {J.~T.}\
  \bibnamefont {Park}}, \bibinfo {author} {\bibfnamefont {T.~G.}\ \bibnamefont
  {Perring}}, \bibinfo {author} {\bibfnamefont {J.~R.}\ \bibnamefont
  {Stewart}}, \bibinfo {author} {\bibfnamefont {H.}~\bibnamefont {Cao}},
  \bibinfo {author} {\bibfnamefont {R.}~\bibnamefont {Zhang}}, \bibinfo
  {author} {\bibfnamefont {Y.}~\bibnamefont {Li}}, \bibinfo {author}
  {\bibfnamefont {Y.}~\bibnamefont {Rong}}, \bibinfo {author} {\bibfnamefont
  {Y.}~\bibnamefont {Wei}}, \bibinfo {author} {\bibfnamefont {B.~M.}\
  \bibnamefont {Andersen}}, \bibinfo {author} {\bibfnamefont {P.~J.}\
  \bibnamefont {Hirschfeld}}, \bibinfo {author} {\bibfnamefont
  {C.}~\bibnamefont {Broholm}}, \ and\ \bibinfo {author} {\bibfnamefont
  {P.}~\bibnamefont {Dai}},\ }\href {\doibase 10.1038/s41563-019-0369-5}
  {\bibfield  {journal} {\bibinfo  {journal} {Nat. Mater.}\ }\textbf {\bibinfo
  {volume} {18}},\ \bibinfo {pages} {709} (\bibinfo {year} {2019})}\BibitemShut
  {NoStop}%
\bibitem [{\citenamefont {Glasbrenner}\ \emph {et~al.}(2015)\citenamefont
  {Glasbrenner}, \citenamefont {Mazin}, \citenamefont {Jeschke}, \citenamefont
  {Hirschfeld}, \citenamefont {Fernandes},\ and\ \citenamefont
  {Valent{\'\i}}}]{Glasbrenner2015Nat.Pys.}%
  \BibitemOpen
  \bibfield  {author} {\bibinfo {author} {\bibfnamefont {J.~K.}\ \bibnamefont
  {Glasbrenner}}, \bibinfo {author} {\bibfnamefont {I.~I.}\ \bibnamefont
  {Mazin}}, \bibinfo {author} {\bibfnamefont {H.~O.}\ \bibnamefont {Jeschke}},
  \bibinfo {author} {\bibfnamefont {P.~J.}\ \bibnamefont {Hirschfeld}},
  \bibinfo {author} {\bibfnamefont {R.~M.}\ \bibnamefont {Fernandes}}, \ and\
  \bibinfo {author} {\bibfnamefont {R.}~\bibnamefont {Valent{\'\i}}},\ }\href
  {https://doi.org/10.1038/nphys3434} {\bibfield  {journal} {\bibinfo
  {journal} {Nat. Phys.}\ }\textbf {\bibinfo {volume} {11}},\ \bibinfo
  {pages} {953} (\bibinfo {year} {2015})}\BibitemShut {NoStop}%
\bibitem [{Note2()}]{Note2}%
  \BibitemOpen
  \bibinfo {note} {There are two inequivalent [1 1 0] directions in the
  spin-polarized DFPT phonon calculation, but the anisotropy is not so
  pronounced between these two direction. Hence, the calculated phonon
  dispersion is shown along one particular [1 1 0] direction}\BibitemShut
  {NoStop}%
\bibitem [{\citenamefont {Bendele}\ \emph {et~al.}(2010)\citenamefont
  {Bendele}, \citenamefont {Amato}, \citenamefont {Conder}, \citenamefont
  {Elender}, \citenamefont {Keller}, \citenamefont {Klauss}, \citenamefont
  {Luetkens}, \citenamefont {Pomjakushina}, \citenamefont {Raselli},\ and\
  \citenamefont {Khasanov}}]{Bendele2010PhysRevLett}%
  \BibitemOpen
  \bibfield  {author} {\bibinfo {author} {\bibfnamefont {M.}~\bibnamefont
  {Bendele}}, \bibinfo {author} {\bibfnamefont {A.}~\bibnamefont {Amato}},
  \bibinfo {author} {\bibfnamefont {K.}~\bibnamefont {Conder}}, \bibinfo
  {author} {\bibfnamefont {M.}~\bibnamefont {Elender}}, \bibinfo {author}
  {\bibfnamefont {H.}~\bibnamefont {Keller}}, \bibinfo {author} {\bibfnamefont
  {H.-H.}\ \bibnamefont {Klauss}}, \bibinfo {author} {\bibfnamefont
  {H.}~\bibnamefont {Luetkens}}, \bibinfo {author} {\bibfnamefont
  {E.}~\bibnamefont {Pomjakushina}}, \bibinfo {author} {\bibfnamefont
  {A.}~\bibnamefont {Raselli}}, \ and\ \bibinfo {author} {\bibfnamefont
  {R.}~\bibnamefont {Khasanov}},\ }\href {\doibase 10.1103/PhysRevLett.104.087003} {\bibfield  {journal} {\bibinfo  {journal}
  {Phys. Rev. Lett.}\ }\textbf {\bibinfo {volume} {104}},\ \bibinfo {pages}
  {087003} (\bibinfo {year} {2010})}\BibitemShut {NoStop}%
\bibitem [{\citenamefont {Bendele}\ \emph {et~al.}(2012)\citenamefont
  {Bendele}, \citenamefont {Ichsanow}, \citenamefont {Pashkevich},
  \citenamefont {Keller}, \citenamefont {Str\"assle}, \citenamefont {Gusev},
  \citenamefont {Pomjakushina}, \citenamefont {Conder}, \citenamefont
  {Khasanov},\ and\ \citenamefont {Keller}}]{Bendele2012Phys.Rev.B}%
  \BibitemOpen
  \bibfield  {author} {\bibinfo {author} {\bibfnamefont {M.}~\bibnamefont
  {Bendele}}, \bibinfo {author} {\bibfnamefont {A.}~\bibnamefont {Ichsanow}},
  \bibinfo {author} {\bibfnamefont {Y.}~\bibnamefont {Pashkevich}}, \bibinfo
  {author} {\bibfnamefont {L.}~\bibnamefont {Keller}}, \bibinfo {author}
  {\bibfnamefont {T.}~\bibnamefont {Str\"assle}}, \bibinfo {author}
  {\bibfnamefont {A.}~\bibnamefont {Gusev}}, \bibinfo {author} {\bibfnamefont
  {E.}~\bibnamefont {Pomjakushina}}, \bibinfo {author} {\bibfnamefont
  {K.}~\bibnamefont {Conder}}, \bibinfo {author} {\bibfnamefont
  {R.}~\bibnamefont {Khasanov}}, \ and\ \bibinfo {author} {\bibfnamefont
  {H.}~\bibnamefont {Keller}},\ }\href {\doibase 10.1103/PhysRevB.85.064517}
  {\bibfield  {journal} {\bibinfo  {journal} {Phys. Rev. B}\ }\textbf {\bibinfo
  {volume} {85}},\ \bibinfo {pages} {064517} (\bibinfo {year}
  {2012})}\BibitemShut {NoStop}%
\bibitem [{\citenamefont {Kothapalli}\ \emph {et~al.}(2016)\citenamefont
  {Kothapalli}, \citenamefont {B{\"o}hmer}, \citenamefont {Jayasekara},
  \citenamefont {Ueland}, \citenamefont {Das}, \citenamefont {Sapkota},
  \citenamefont {Taufour}, \citenamefont {Xiao}, \citenamefont {Alp},
  \citenamefont {Bud'ko}, \citenamefont {Canfield}, \citenamefont {Kreyssig},\
  and\ \citenamefont {Goldman}}]{Kothapalli2016Nat.Commun}%
  \BibitemOpen
  \bibfield  {author} {\bibinfo {author} {\bibfnamefont {K.}~\bibnamefont
  {Kothapalli}}, \bibinfo {author} {\bibfnamefont {A.~E.}\ \bibnamefont
  {B{\"o}hmer}}, \bibinfo {author} {\bibfnamefont {W.~T.}\ \bibnamefont
  {Jayasekara}}, \bibinfo {author} {\bibfnamefont {B.~G.}\ \bibnamefont
  {Ueland}}, \bibinfo {author} {\bibfnamefont {P.}~\bibnamefont {Das}},
  \bibinfo {author} {\bibfnamefont {A.}~\bibnamefont {Sapkota}}, \bibinfo
  {author} {\bibfnamefont {V.}~\bibnamefont {Taufour}}, \bibinfo {author}
  {\bibfnamefont {Y.}~\bibnamefont {Xiao}}, \bibinfo {author} {\bibfnamefont
  {E.}~\bibnamefont {Alp}}, \bibinfo {author} {\bibfnamefont {S.~L.}\
  \bibnamefont {Bud'ko}}, \bibinfo {author} {\bibfnamefont {P.~C.}\
  \bibnamefont {Canfield}}, \bibinfo {author} {\bibfnamefont {A.}~\bibnamefont
  {Kreyssig}}, \ and\ \bibinfo {author} {\bibfnamefont {A.~I.}\ \bibnamefont
  {Goldman}},\ }\href {\doibase 10.1038/ncomms12728} {\bibfield  {journal}
  {\bibinfo  {journal} {Nat. Commun}\ }\textbf {\bibinfo {volume} {7}},\
  \bibinfo {pages} {12728} (\bibinfo {year} {2016})}\BibitemShut {NoStop}%
\bibitem [{\citenamefont {B\"ohmer}\ \emph {et~al.}(2019)\citenamefont
  {B\"ohmer}, \citenamefont {Kothapalli}, \citenamefont {Jayasekara},
  \citenamefont {Wilde}, \citenamefont {Li}, \citenamefont {Sapkota},
  \citenamefont {Ueland}, \citenamefont {Das}, \citenamefont {Xiao},
  \citenamefont {Bi}, \citenamefont {Zhao}, \citenamefont {Alp}, \citenamefont
  {Bud'ko}, \citenamefont {Canfield}, \citenamefont {Goldman},\ and\
  \citenamefont {Kreyssig}}]{Bohmer2019PhysRevB}%
  \BibitemOpen
  \bibfield  {author} {\bibinfo {author} {\bibfnamefont {A.~E.}\ \bibnamefont
  {B\"ohmer}}, \bibinfo {author} {\bibfnamefont {K.}~\bibnamefont
  {Kothapalli}}, \bibinfo {author} {\bibfnamefont {W.~T.}\ \bibnamefont
  {Jayasekara}}, \bibinfo {author} {\bibfnamefont {J.~M.}\ \bibnamefont
  {Wilde}}, \bibinfo {author} {\bibfnamefont {B.}~\bibnamefont {Li}}, \bibinfo
  {author} {\bibfnamefont {A.}~\bibnamefont {Sapkota}}, \bibinfo {author}
  {\bibfnamefont {B.~G.}\ \bibnamefont {Ueland}}, \bibinfo {author}
  {\bibfnamefont {P.}~\bibnamefont {Das}}, \bibinfo {author} {\bibfnamefont
  {Y.}~\bibnamefont {Xiao}}, \bibinfo {author} {\bibfnamefont {W.}~\bibnamefont
  {Bi}}, \bibinfo {author} {\bibfnamefont {J.}~\bibnamefont {Zhao}}, \bibinfo
  {author} {\bibfnamefont {E.~E.}\ \bibnamefont {Alp}}, \bibinfo {author}
  {\bibfnamefont {S.~L.}\ \bibnamefont {Bud'ko}}, \bibinfo {author}
  {\bibfnamefont {P.~C.}\ \bibnamefont {Canfield}}, \bibinfo {author}
  {\bibfnamefont {A.~I.}\ \bibnamefont {Goldman}}, \ and\ \bibinfo {author}
  {\bibfnamefont {A.}~\bibnamefont {Kreyssig}},\ }\href {\doibase 10.1103/PhysRevB.100.064515} {\bibfield  {journal} {\bibinfo  {journal}
  {Phys. Rev. B}\ }\textbf {\bibinfo {volume} {100}},\ \bibinfo {pages}
  {064515} (\bibinfo {year} {2019})}\BibitemShut {NoStop}%
\bibitem [{\citenamefont {K\"ormann}\ \emph {et~al.}(2012)\citenamefont
  {K\"ormann}, \citenamefont {Dick}, \citenamefont {Grabowski}, \citenamefont
  {Hickel},\ and\ \citenamefont {Neugebauer}}]{Kormann2012Phys.Rev.B}%
  \BibitemOpen
  \bibfield  {author} {\bibinfo {author} {\bibfnamefont {F.}~\bibnamefont
  {K\"ormann}}, \bibinfo {author} {\bibfnamefont {A.}~\bibnamefont {Dick}},
  \bibinfo {author} {\bibfnamefont {B.}~\bibnamefont {Grabowski}}, \bibinfo
  {author} {\bibfnamefont {T.}~\bibnamefont {Hickel}}, \ and\ \bibinfo {author}
  {\bibfnamefont {J.}~\bibnamefont {Neugebauer}},\ }\href {\doibase 10.1103/PhysRevB.85.125104} {\bibfield  {journal} {\bibinfo  {journal} {Phys.
  Rev. B}\ }\textbf {\bibinfo {volume} {85}},\ \bibinfo {pages} {125104}
  (\bibinfo {year} {2012})}\BibitemShut {NoStop}%
\bibitem [{\citenamefont {Krannich}\ \emph {et~al.}(2015)\citenamefont
  {Krannich}, \citenamefont {Sidis}, \citenamefont {Lamago}, \citenamefont
  {Heid}, \citenamefont {Mignot}, \citenamefont {L{\"o}hneysen}, \citenamefont
  {Ivanov}, \citenamefont {Steffens}, \citenamefont {Keller}, \citenamefont
  {Wang}, \citenamefont {Goering},\ and\ \citenamefont
  {Weber}}]{Krannich2015Nat.Commun.}%
  \BibitemOpen
  \bibfield  {author} {\bibinfo {author} {\bibfnamefont {S.}~\bibnamefont
  {Krannich}}, \bibinfo {author} {\bibfnamefont {Y.}~\bibnamefont {Sidis}},
  \bibinfo {author} {\bibfnamefont {D.}~\bibnamefont {Lamago}}, \bibinfo
  {author} {\bibfnamefont {R.}~\bibnamefont {Heid}}, \bibinfo {author}
  {\bibfnamefont {J.~M.}\ \bibnamefont {Mignot}}, \bibinfo {author}
  {\bibfnamefont {H.~v.}\ \bibnamefont {L{\"o}hneysen}}, \bibinfo {author}
  {\bibfnamefont {A.}~\bibnamefont {Ivanov}}, \bibinfo {author} {\bibfnamefont
  {P.}~\bibnamefont {Steffens}}, \bibinfo {author} {\bibfnamefont
  {T.}~\bibnamefont {Keller}}, \bibinfo {author} {\bibfnamefont
  {L.}~\bibnamefont {Wang}}, \bibinfo {author} {\bibfnamefont {E.}~\bibnamefont
  {Goering}}, \ and\ \bibinfo {author} {\bibfnamefont {F.}~\bibnamefont
  {Weber}},\ }\href {https://doi.org/10.1038/ncomms9961} {\bibfield  {journal}
  {\bibinfo  {journal} {Nat. Commun.}\ }\textbf {\bibinfo {volume} {6}},\
  \bibinfo {pages} {8961} (\bibinfo {year} {2015})}\BibitemShut {NoStop}%
\bibitem [{\citenamefont {Han}\ \emph {et~al.}(2018)\citenamefont {Han},
  \citenamefont {Birol},\ and\ \citenamefont {Haule}}]{Han2018Phys.Rev.Lett}%
  \BibitemOpen
  \bibfield  {author} {\bibinfo {author} {\bibfnamefont {Q.}~\bibnamefont
  {Han}}, \bibinfo {author} {\bibfnamefont {T.}~\bibnamefont {Birol}}, \ and\
  \bibinfo {author} {\bibfnamefont {K.}~\bibnamefont {Haule}},\ }\href
  {\doibase 10.1103/PhysRevLett.120.187203} {\bibfield  {journal} {\bibinfo
  {journal} {Phys. Rev. Lett.}\ }\textbf {\bibinfo {volume} {120}},\ \bibinfo
  {pages} {187203} (\bibinfo {year} {2018})}\BibitemShut {NoStop}%
\bibitem [{\citenamefont {Boeri}\ \emph {et~al.}(2010)\citenamefont {Boeri},
  \citenamefont {Calandra}, \citenamefont {Mazin}, \citenamefont {Dolgov},\
  and\ \citenamefont {Mauri}}]{Boeri2010Phys.RevB.}%
  \BibitemOpen
  \bibfield  {author} {\bibinfo {author} {\bibfnamefont {L.}~\bibnamefont
  {Boeri}}, \bibinfo {author} {\bibfnamefont {M.}~\bibnamefont {Calandra}},
  \bibinfo {author} {\bibfnamefont {I.~I.}\ \bibnamefont {Mazin}}, \bibinfo
  {author} {\bibfnamefont {O.~V.}\ \bibnamefont {Dolgov}}, \ and\ \bibinfo
  {author} {\bibfnamefont {F.}~\bibnamefont {Mauri}},\ }\href {\doibase 10.1103/PhysRevB.82.020506} {\bibfield  {journal} {\bibinfo  {journal} {Phys.
  Rev. B}\ }\textbf {\bibinfo {volume} {82}},\ \bibinfo {pages} {020506}
  (\bibinfo {year} {2010})}\BibitemShut {NoStop}%
\bibitem [{\citenamefont {Coh}\ \emph {et~al.}(2015)\citenamefont {Coh},
  \citenamefont {Cohen},\ and\ \citenamefont {Louie}}]{Coh2015New.J.Phys.}%
  \BibitemOpen
  \bibfield  {author} {\bibinfo {author} {\bibfnamefont {S.}~\bibnamefont
  {Coh}}, \bibinfo {author} {\bibfnamefont {M.~L.}\ \bibnamefont {Cohen}}, \
  and\ \bibinfo {author} {\bibfnamefont {S.~G.}\ \bibnamefont {Louie}},\ }\href
  {\doibase 10.1088/1367-2630/17/7/073027} {\bibfield  {journal} {\bibinfo
  {journal} {New J. Phys.}\ }\textbf {\bibinfo {volume} {17}},\ \bibinfo
  {pages} {073027} (\bibinfo {year} {2015})}\BibitemShut {NoStop}%
\bibitem [{\citenamefont {Coh}\ \emph {et~al.}(2016)\citenamefont {Coh},
  \citenamefont {Cohen},\ and\ \citenamefont {Louie}}]{Coh2016Phys.Rev.B}%
  \BibitemOpen
  \bibfield  {author} {\bibinfo {author} {\bibfnamefont {S.}~\bibnamefont
  {Coh}}, \bibinfo {author} {\bibfnamefont {M.~L.}\ \bibnamefont {Cohen}}, \
  and\ \bibinfo {author} {\bibfnamefont {S.~G.}\ \bibnamefont {Louie}},\ }\href
  {\doibase 10.1103/PhysRevB.94.104505} {\bibfield  {journal} {\bibinfo
  {journal} {Phys. Rev. B}\ }\textbf {\bibinfo {volume} {94}},\ \bibinfo
  {pages} {104505} (\bibinfo {year} {2016})}\BibitemShut {NoStop}%
\bibitem [{\citenamefont {Mandal}\ \emph {et~al.}(2014)\citenamefont {Mandal},
  \citenamefont {Cohen},\ and\ \citenamefont {Haule}}]{Mandal2014Phys.Rev.B}%
  \BibitemOpen
  \bibfield  {author} {\bibinfo {author} {\bibfnamefont {S.}~\bibnamefont
  {Mandal}}, \bibinfo {author} {\bibfnamefont {R.~E.}\ \bibnamefont {Cohen}}, \
  and\ \bibinfo {author} {\bibfnamefont {K.}~\bibnamefont {Haule}},\ }\href
  {\doibase 10.1103/PhysRevB.89.220502} {\bibfield  {journal} {\bibinfo
  {journal} {Phys. Rev. B}\ }\textbf {\bibinfo {volume} {89}},\ \bibinfo
  {pages} {220502} (\bibinfo {year} {2014})}\BibitemShut {NoStop}%
\bibitem [{\citenamefont {Gerber}\ \emph {et~al.}(2017)\citenamefont {Gerber},
  \citenamefont {Yang}, \citenamefont {Zhu}, \citenamefont {Soifer},
  \citenamefont {Sobota}, \citenamefont {Rebec}, \citenamefont {Lee},
  \citenamefont {Jia}, \citenamefont {Moritz}, \citenamefont {Jia},
  \citenamefont {Gauthier}, \citenamefont {Li}, \citenamefont {Leuenberger},
  \citenamefont {Zhang}, \citenamefont {Chaix}, \citenamefont {Li},
  \citenamefont {Jang}, \citenamefont {Lee}, \citenamefont {Yi}, \citenamefont
  {Dakovski}, \citenamefont {Song}, \citenamefont {Glownia}, \citenamefont
  {Nelson}, \citenamefont {Kim}, \citenamefont {Chuang}, \citenamefont
  {Hussain}, \citenamefont {Moore}, \citenamefont {Devereaux}, \citenamefont
  {Lee}, \citenamefont {Kirchmann},\ and\ \citenamefont
  {Shen}}]{Gerber2017Science}%
  \BibitemOpen
  \bibfield  {author} {\bibinfo {author} {\bibfnamefont {S.}~\bibnamefont
  {Gerber}}, \bibinfo {author} {\bibfnamefont {S.-L.}\ \bibnamefont {Yang}},
  \bibinfo {author} {\bibfnamefont {D.}~\bibnamefont {Zhu}}, \bibinfo {author}
  {\bibfnamefont {H.}~\bibnamefont {Soifer}}, \bibinfo {author} {\bibfnamefont
  {J.~A.}\ \bibnamefont {Sobota}}, \bibinfo {author} {\bibfnamefont
  {S.}~\bibnamefont {Rebec}}, \bibinfo {author} {\bibfnamefont {J.~J.}\
  \bibnamefont {Lee}}, \bibinfo {author} {\bibfnamefont {T.}~\bibnamefont
  {Jia}}, \bibinfo {author} {\bibfnamefont {B.}~\bibnamefont {Moritz}},
  \bibinfo {author} {\bibfnamefont {C.}~\bibnamefont {Jia}}, \bibinfo {author}
  {\bibfnamefont {A.}~\bibnamefont {Gauthier}}, \bibinfo {author}
  {\bibfnamefont {Y.}~\bibnamefont {Li}}, \bibinfo {author} {\bibfnamefont
  {D.}~\bibnamefont {Leuenberger}}, \bibinfo {author} {\bibfnamefont
  {Y.}~\bibnamefont {Zhang}}, \bibinfo {author} {\bibfnamefont
  {L.}~\bibnamefont {Chaix}}, \bibinfo {author} {\bibfnamefont
  {W.}~\bibnamefont {Li}}, \bibinfo {author} {\bibfnamefont {H.}~\bibnamefont
  {Jang}}, \bibinfo {author} {\bibfnamefont {J.-S.}\ \bibnamefont {Lee}},
  \bibinfo {author} {\bibfnamefont {M.}~\bibnamefont {Yi}}, \bibinfo {author}
  {\bibfnamefont {G.~L.}\ \bibnamefont {Dakovski}}, \bibinfo {author}
  {\bibfnamefont {S.}~\bibnamefont {Song}}, \bibinfo {author} {\bibfnamefont
  {J.~M.}\ \bibnamefont {Glownia}}, \bibinfo {author} {\bibfnamefont
  {S.}~\bibnamefont {Nelson}}, \bibinfo {author} {\bibfnamefont {K.~W.}\
  \bibnamefont {Kim}}, \bibinfo {author} {\bibfnamefont {Y.-D.}\ \bibnamefont
  {Chuang}}, \bibinfo {author} {\bibfnamefont {Z.}~\bibnamefont {Hussain}},
  \bibinfo {author} {\bibfnamefont {R.~G.}\ \bibnamefont {Moore}}, \bibinfo
  {author} {\bibfnamefont {T.~P.}\ \bibnamefont {Devereaux}}, \bibinfo {author}
  {\bibfnamefont {W.-S.}\ \bibnamefont {Lee}}, \bibinfo {author} {\bibfnamefont
  {P.~S.}\ \bibnamefont {Kirchmann}}, \ and\ \bibinfo {author} {\bibfnamefont
  {Z.-X.}\ \bibnamefont {Shen}},\ }\href {\doibase 10.1126/science.aak9946}
  {\bibfield  {journal} {\bibinfo  {journal} {Science}\ }\textbf {\bibinfo
  {volume} {357}},\ \bibinfo {pages} {71} (\bibinfo {year} {2017})}\BibitemShut
  {NoStop}%
\end{thebibliography}%
\section{APPENDIX: Structural vs magnetic contributions to the phonon dispersion}
\label{appendix}
\begin{figure*}[htb]
\begin{center}
  \includegraphics[clip,width= 18.0000cm]{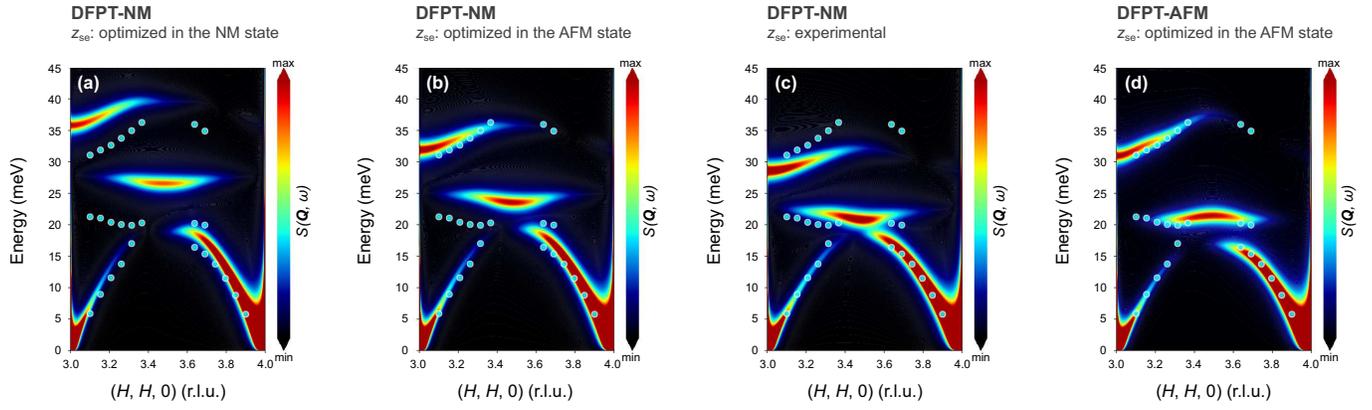}
  \caption{
    (Color online) Comparison of the measured phonon dispersion for FeSe and DFPT calculations in the
    NM ((a)-(c)) and the AFM (d) states.
    To understand the structural effects on phonon, non-magnetic DFPT calculations ((a)-(c)) were performed
    using different values of {\it z$_{{\rm { }Se}}$}.
    In (a) and (b), phonon calculations were performed using the values of {\it z$_{{\rm { }Se}}$} optimized in the
    NM and AFM states, respectively, whereas in (c), phonon calculation was performed
    using the experimental {\it z$_{{\rm { }Se}}$} 
    without structural optimization. 
  }
\label{Fig4}
\end{center}
\end{figure*}
\indent Phonon calculations presented in the main text were performed using the DFT-optimized
crystal structure, and thus the resulting phonon dispersion is affected not only by the
magnetic ground state but also by the structural details. To disentangle the structural and
magnetic contributions to the phonon dispersion, we performed the non-magnetic DFPT 
calculation using the value of {\it z$_{ \scalebox{0.55}{\rm { }Se}}$} optimized in the AFM state. 
As can be seen in Figs. \ref{Fig4} (a) and (b), this calculation gives better overall agreement with the
experiment than that obtained by using {\it z$_{ \scalebox{0.55}{\rm { }Se}}$} optimized in the NM state. 
This improvement is due to a better description of {\it z$_{ \scalebox{0.55}{\rm { }Se}}$} in the
spin-polarized DFT calculation. Further improvement can be obtained by using the experimental
{\it z$_{ \scalebox{0.55}{\rm { }Se}}$} (see Fig. \ref{Fig4} (c)). There are, however, still some discrepancies between
the calculated and measured phonon dispersion.
A good quantitative description of the experimental data can only be achieved
by imposing the AFM order in the DFT calculation. 
(see Fig. \ref{Fig4} (d)). 
Therefore, the structural parameters alone are not sufficient for the description of the experimental phonon dispersion of FeSe,
and the inclusion of magnetism is crucial. 
\end{document}